\documentclass[aps,pra,twocolumn,superscriptaddress]{revtex4-1}

\usepackage{amsmath,amssymb,amsfonts}
\usepackage{graphicx}
\usepackage{siunitx}
\usepackage{natbib}
\usepackage{braket}
\usepackage{multirow}

\usepackage{url}
\usepackage[breaklinks]{hyperref}
\hypersetup{
    unicode=true,
    colorlinks=true,
    linkcolor=blue,
    citecolor=blue,
    urlcolor=blue
  }
	\usepackage{breakurl}

\graphicspath{ {Figures/} }

\begin{document}

\title{Two-photon photoassociation spectroscopy of CsYb:\\
ground-state interaction potential and interspecies scattering lengths}

\author{A. Guttridge}
 \affiliation{Joint Quantum Centre (JQC) Durham-Newcastle, Department of Physics, Durham University, South Road, Durham, DH1 3LE, United Kingdom.}
\author{Matthew D. Frye}
\affiliation{Joint Quantum Centre (JQC) Durham-Newcastle, Department of Chemistry, Durham University, South Road, Durham, DH1 3LE, United Kingdom.}
\author{B. C. Yang}
\affiliation{Joint Quantum Centre (JQC) Durham-Newcastle, Department of Chemistry, Durham University, South Road, Durham, DH1 3LE, United Kingdom.}
\author{Jeremy M. Hutson}
\email{j.m.hutson@durham.ac.uk}
\affiliation{Joint Quantum Centre (JQC) Durham-Newcastle, Department of Chemistry, Durham University, South Road, Durham, DH1 3LE, United Kingdom.}
\author{Simon L. Cornish}
\email{s.l.cornish@durham.ac.uk}
\affiliation{Joint Quantum Centre (JQC) Durham-Newcastle, Department of Physics, Durham University, South Road, Durham, DH1 3LE, United Kingdom.}

\begin{abstract}
We perform two-photon photoassociation spectroscopy of the heteronuclear CsYb
molecule to measure the binding energies of near-threshold vibrational levels
of the $X~^{2}\Sigma_{1/2}^{+}$ molecular ground state. We report results for
$^{133}$Cs$^{170}$Yb, $^{133}$Cs$^{173}$Yb and $^{133}$Cs$^{174}$Yb, in each
case determining the energy of several vibrational levels including the
least-bound state. We fit an interaction potential based on electronic structure
calculations to the binding energies for all three isotopologs and find that
the ground-state potential supports 77 vibrational levels. We use the fitted
potential to predict the interspecies s-wave scattering lengths for all seven
Cs+Yb isotopic mixtures.
\end{abstract}

\date{\today}

\maketitle

\section{Introduction}

Mixtures of ultracold atomic gases provide an appealing platform for numerous
avenues of research, including the investigation of novel quantum phases
\cite{Molmer1998,Lewenstein2004a,Zaccanti2006a,Ospelkaus2006a,Guenter2006,Sengupta2007,Marchetti2008},
the study of Efimov physics \cite{Tung2014,Pires2014,Maier2015,Ulmanis2016a}
and the creation of ultracold polar molecules
\cite{Koehler2006,Ni2008,Lang2008,Aikawa2009,Koeppinger2014,Molony2014,Molony2016,Takekoshi2014,Park2015,Guo2016,Bohn2017}.
Early experiments explored bi-alkali-metal gases
\cite{Modugno2001,Mudrich2002,Hadzibabic2002,Taglieber2008,Spiegelhalder2009,
Taie2010,Cho2011,McCarron2011,Ridinger2011,Wacker2015,Grobner2016}, but there
is currently a growing interest in mixtures composed of alkali-metal and
closed-shell atoms
\cite{Tassy2010,Hara2014,Pasquiou2013,Khramov2014,Vaidya2015,Guttridge2017,Flores2017,Witkowski2017}.
Such mixtures open up the possibility of creating paramagnetic ground-state
polar molecules, with applications in quantum simulation and quantum
information \cite{Micheli2006,Perez-Rios2010,Herrera2014}, precision
measurement \cite{Alyabyshev2012}, tests of fundamental physics
\cite{Isaev2010,Flambaum2007,Hudson2011} and tuning of collisions and chemical
reactions \cite{Abrahamsson2007,Quemener2016}. In pursuit of this goal we have
constructed an apparatus to investigate ultracold mixtures of Cs and Yb
\cite{Hopkins2016,Kemp2016,Guttridge2016}.

Magnetoassociation on a Feshbach resonance has proved a highly successful
technique for producing weakly bound ultracold molecules
\cite{Koehler2006,Chin2010}. When combined with optical transfer using
Stimulated Raman Adiabatic Passage (STIRAP), the approach has allowed the
production of a range of ultracold polar bi-alkali molecules in the
rovibrational ground state
\cite{Ni2008,Takekoshi2014,Molony2014,Park2015,Guo2016}. Unfortunately, in the
case of an alkali-metal atom and a closed-shell atom, the Feshbach resonances
are predicted to be narrow and sparsely distributed in magnetic field
\cite{Zuchowski2010,Brue2012}. Nevertheless, such resonances have recently been
observed experimentally in the RbSr system \cite{Barbe2017}, though
magnetoassociation remains unexplored. The resonances in CsYb are predicted to
be particularly favorable for magnetoassociation \cite{Brue2013}. However, to
predict their locations accurately it is necessary first to determine the
binding energies of the near-threshold vibrational levels of the CsYb molecule.

In this paper we present two-photon photoassociation spectroscopy of the
heteronuclear CsYb molecule. Using ultracold mixtures of Cs and Yb confined in
an optical dipole trap, we accurately measure the binding energies of the
near-threshold vibrational levels of CsYb molecules in the
$X~^{2}\Sigma_{1/2}^{+}$ ground state. We report results for three isotopologs,
$^{133}$Cs$^{170}$Yb, $^{133}$Cs$^{173}$Yb and $^{133}$Cs$^{174}$Yb, in each
case measuring the energy of several vibrational levels including the
least-bound state. We fit an interaction potential based on electronic
structure calculations to the binding energies for all three isotopologs and
find that the ground-state potential supports 77 vibrational levels. The
excellent agreement between our model and the experimental results allows us to
calculate the interspecies scattering lengths for $^{133}$Cs interacting with
all seven stable Yb isotopes.

\section{Two-photon Photoassociation spectroscopy}
\begin{figure}
		\includegraphics[width=0.95\linewidth]{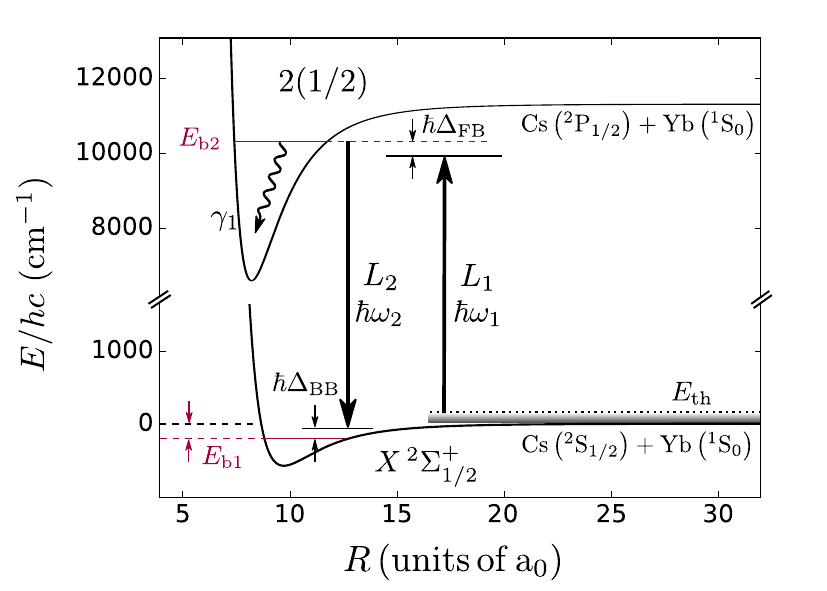}
\caption{Two-photon photoassociation for the measurement of the
binding energy, $E_{\rm b1}$, of a vibrational level of CsYb in
its electronic ground state. A pair of colliding Cs and Yb atoms with thermal
energy $E_{\rm th}$ is associated to form a CsYb molecule in a rovibrational
level of the electronically excited $2(1/2)$ state by light of
frequency $\omega_{1}$. This rovibrational level is coupled to a level in the electronic
ground state, $X \, ^{2}\Sigma^{+}_{1/2}$, by light of frequency $\omega_{2}$. The
molecular curves plotted here are adapted from Ref.\ \cite{Meniailava2017}. The
internuclear distances where the transitions occur are not shown to scale.
	\label{fig:2PA figure}}
\end{figure}

\subsection{Overview}

The two-photon photoassociation process is shown in Fig.\ \ref{fig:2PA figure}.
This scheme is an extension of one-photon photoassociation
\cite{Bohn1999,Jones2006}, whereby a pair of colliding atoms is associated to
form a molecule in a rovibrational level of an electronically excited molecular
state. The laser that drives the one-photon photoassociation, $L_{1}$, has
frequency $\omega_{1}$ and intensity $I_1$; it is detuned from a free-bound
transition by $\Delta_{\rm FB}$. The second laser, $L_{2}$, has frequency
$\omega_{2}$ and intensity $I_2$; it couples the electronically excited
molecule to a rovibrational level of the molecule in the electronic ground
state. Its detuning from this bound-bound transition is $\Delta_{\rm BB}$. When
$L_{2}$ is resonant with a bound-bound transition, the coupling leads to the
formation of a dark state and the suppression of the absorption of $L_1$. Such
two-photon dark resonances can be used to measure the binding energies, $E_{\rm
b1}$, of vibrational levels of the molecule in the electronic ground state. In
the undressed, zero-temperature limit, the binding energy is given simply by
the difference in photon energy of the two lasers, $E_{\rm b1} = \hbar
\left(\omega_{2} - \omega_{1}\right)$, when on two-photon resonance. This
technique has been applied in a large number of single-species
\cite{Abraham1996,Tsai1997,Abeelen1999a,Wang2000,
Vanhaecke2004,Moal2006,MartinezdeEscobar2008,Kitagawa2008,Gunton2013,Pachomow2017}
and two-species ultracold atom experiments
\cite{Ni2008,Munchow2011,Debatin2011,Guo2017,Dutta2017,Rvachov2018} with
considerable success.

For the specific case of CsYb discussed in this paper, the first photon excites
the colliding atoms into a rovibrational level of the molecule close to the
Cs($^{2}P_{1/2}$) + Yb($^{1}S_{0}$) asymptote. The electronic state at this
threshold is designated 2(1/2) to indicate that it is the second (first
excited) state with total electronic angular momentum $\Omega=1/2$ about the
internuclear axis. It correlates at short range with the $1\,^2\Pi_{1/2}$
electronic state in Hund's case (a) notation \cite{Meniailava2017}, but at long
range the $1 \, ^2\Pi_{1/2}$ and $2 \, ^2\Sigma_{1/2}$ states are strongly mixed by
spin-orbit coupling. We have recently reported photoassociation spectroscopy of
the vibrational levels of the molecule within 500\,GHz of the 2(1/2) threshold
\cite{Guttridge2018}. In this work we add a second photon to couple the
vibrational level in the electronically excited state to a near-threshold level
of the $X \, ^{2}\Sigma^{+}_{1/2}$ electronic ground state. We label each
vibrational level by its vibrational number $n$ below the associated threshold,
such that $n=-1$ corresponds to the least-bound state, using $n'$ for the
electronically excited state and $n''$ for the ground state. Because of the low
temperature of our atomic mixtures, combined with the selection rule $\Delta
N=0$, all the rovibrational levels we measure have rotational quantum number
$N=0$.

\subsection{Experimental Setup}

The experimental setup has been described in the context of our previous work
\cite{Kemp2016,Hopkins2016,Guttridge2016,Guttridge2017,Guttridge2018}. Here we
focus on details of the ultracold atomic mixtures and the two-photon
photoassociation setup.

Our measurements are performed on mixtures of Cs and Yb confined in an optical
dipole trap (ODT). The ODT is formed from the output of a broadband fiber laser
(IPG YLR-100-LP) with a wavelength of $1070(3) \,$nm and consists of two beams
crossed at an angle of $40 ^{\circ}$ with waists of $33(4) \, \mu $m and $72(4)
\, \mu$m. The measured Yb(Cs) trap frequencies are $240 (750) \,$Hz radially
and $40 (120)\,$Hz axially. The trap depths for the two species are
$U_{\mathrm{Yb}} = 5 \, \mu$K and $U_{\mathrm{Cs}} = 85 \, \mu$K respectively.
We load the ODT with a mixture of $7 \times 10^{4}$ Cs atoms at
$T_{\mathrm{Cs}} = 6 \, \mu$K in the absolute ground state $6^{2}S_{1/2}$
$\ket{F=3, m_{F}=+3}$ and Yb atoms at $T_{\mathrm{Yb}} = 1 \, \mu$K in the
$^{1}S_{0}$ ground state. The number of Yb atoms depends on the Yb isotope
involved. Typically, we use $8 \times 10^{5}$ atoms for $^{174}$Yb, $4 \times
10^{5}$ atoms for $^{170}$Yb or $3 \times 10^{5}$ atoms for $^{173}$Yb. For
both atomic species the atom number is measured using resonant absorption
imaging after a short time of flight.

The light for two-photon photoassociation is derived from two independent
lasers. $L_1$ is a Ti:Sapphire laser (\mbox{M Squared SolsTiS}) and $L_2$ is a
Distributed Bragg Reflector (DBR) laser. Both lasers are frequency-stabilized
using a high-finesse optical cavity, the length of which is stabilized to a Cs
atomic transition using the Pound-Drever-Hall method \cite{Drever1983}. The
light sent to the optical cavity from $L_1$ and $L_2$ is first passed through
two independent broadband fiber electro-optic modulators (EOMs) (EOSPACE
PM-0S5-10-PFA-PFA-895) to add frequency sidebands. We then utilize the
`electronic sideband' technique \cite{Thorpe2008,Gregory2015} to allow
continuous tuning of the two laser frequencies; by stabilizing a frequency
sideband to a cavity transmission peak, the carrier frequencies of both lasers
may be tuned over the $748.852(5) \,$MHz free spectral range (FSR) of the
cavity by changing the modulation frequencies applied to the EOMs. By
stabilizing the two lasers to different modes of the cavity we can control
their frequency difference, $\omega_{1} - \omega_{2}$, over many GHz.

The main outputs of lasers $L_1$ and $L_2$ are overlapped, transmitted through
an acousto-optic modulator for fast intensity control and coupled into a fiber
that carries the light to the experiment. The output of the fiber is focused
onto the atomic mixture with a waist of $150 \, \mu$m and is circularly
polarized to drive $\sigma^{+}$ transitions. This polarization gives us the
strongest two-photon transitions from the Cs($6 ^{2}S_{1/2} F=3, m_F=+3$) +
Yb($^{1}S_{0}$) scattering state to the $F=3$ manifold of the molecular
electronic ground state via an intermediate vibrational level of CsYb in the
$F'=4$ manifold of the $2(1/2)$ excited state \cite{Guttridge2018}.

We measure the frequency difference between lasers $L_1$ and $L_2$ using one of three methods, depending on the binding energy of the state under investigation. Most generally, the frequency difference is determined from the difference in the modulation frequencies applied to the two EOMs, combined with the number of cavity FSRs between the two modes used for frequency stabilization. Light from both lasers is coupled into a commercial wavemeter (Bristol 671A) for absolute frequency calibration and unambiguous determination of the cavity mode. For binding energies below 2\,GHz, the frequency difference between the two lasers is measured directly from the beat frequency recorded on a fast photodiode (EOT ET-2030A). In the special case of the least-bound state, we do not use DBR laser and instead we drive the AOM with two RF frequencies. Generating the two-photon detuning in this way eliminates any effects of laser frequency noise and allows a very precise determination of the frequency difference.

\subsection{Experimental Results}
\begin{figure}
		\includegraphics[width=0.95\linewidth]{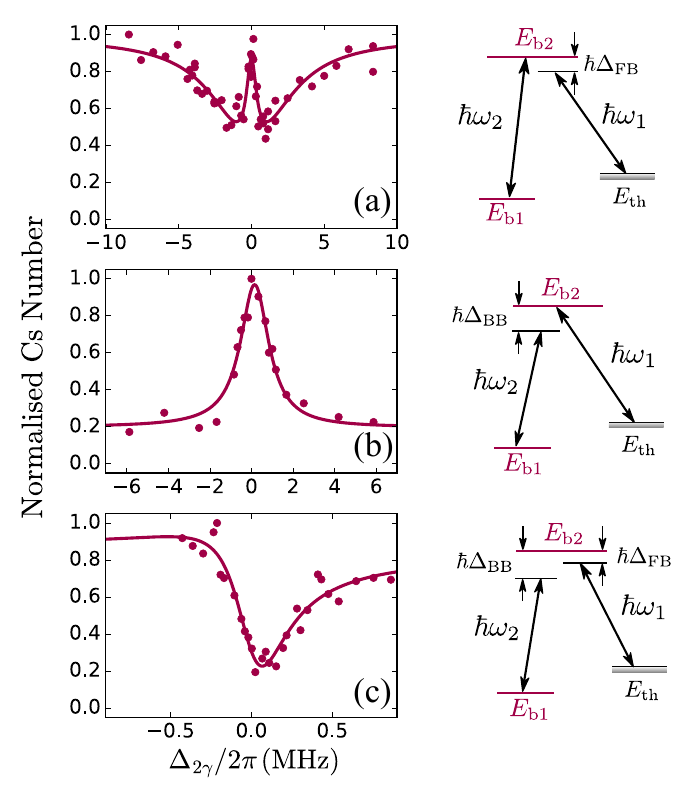}

\caption{Measurement of the least-bound state of Cs$^{174}$Yb in the $X \,
^{2}\Sigma^{+}_{1/2}$ electronic ground state by two-photon photoassociation
spectroscopy. The intermediate state used is the $n' = -13 $ level of the
molecule in the $2(1/2)$ state. Left panel: Two-photon photoassociation
spectra. Normalized number of Cs atoms plotted against $\Delta_{2 \gamma} =
\Delta_{\mathrm{FB}} - \Delta_{\mathrm{BB}}$. Right panel: Simplified level
structure for the two-photon photoassociation transitions. (a) Dark-resonance
spectroscopy performed by scanning $\omega_{1}$. The red solid line shows a fit
using Eq.\ \ref{eq:three-level}, the analytical solution of the optical Bloch
equations for a lambda-type three-level system. The best-fit parameters are
$\Omega_{\rm BB} / 2\pi = 2(1)$~MHz, $\Gamma / 2\pi = 6(1)$~MHz and
$\Gamma_{\rm eff} / 2\pi = 2(1) \times 10^{2}$~kHz. (b) Dark-resonance
spectroscopy performed by scanning $\omega_{2}$. The red solid line shows a fit
using a Lorentzian profile. (c) Raman spectroscopy. The red solid line shows a
fit using a Fano profile. The spectra shown in (a),(b),(c) were obtained with
laser intensities $I_{1} = 0.42, 0.68, 1.1 \, {\rm W/cm^{2}} $ and $I_{2} =
0.79, 0.57, 0.76 \, {\rm W/cm^{2}}$, respectively.
	\label{fig:2PA spectrum}}
\end{figure}

Two-photon photoassociation measurements are performed by illuminating the
atomic mixture with light from lasers $L_{1}$ and $L_{2}$ for a variable time
up to $250 \,$ms, in a magnetic field of \mbox{2.2(2) G}. Figure \ref{fig:2PA
spectrum} shows the two-photon feature for the least-bound $n''=-1$ level of
Cs$^{174}$Yb, using the $n' = -13$ level of the 2(1/2) state as the
intermediate state. We detect the two-photon resonance by measuring the number
of Cs atoms remaining after exposure to the photoassociation light as a
function of the two-photon detuning $\Delta_{2\gamma}= \Delta_{\rm FB} -
\Delta_{\mathrm{BB}}$. Three different lineshapes may be observed, depending on
the relative intensities and detunings of the lasers.

Figure \ref{fig:2PA spectrum}(a) shows the lineshape observed using two-photon
dark-resonance spectroscopy \cite{Winkler2005,Debatin2011}. In this method
$\omega_{2}$ is fixed on resonance with the bound-bound transition
($\Delta_{\mathrm{BB}} = 0$) and $\omega_{1}$ is scanned over the free-bound
transition. The spectrum exhibits the w-shaped profile expected for
electromagnetically induced transparency (EIT) in a lambda-type three-level
system \cite{Fleischhauer2005} and we therefore refer to this as the EIT
lineshape. In the wings we observe a Lorentzian profile originating from
one-photon photoassociation to the $n'=-13$ level of the 2(1/2) state. Then, on
resonance we see a suppression of the photoassociative loss due to the creation
of a dark state composed of the initial atomic scattering state and the
molecular ground state. This dark state is decoupled from the intermediate
$n'=-13$ state and leads to the observed `transparency'. We fit the data with
the analytical solution of the optical Bloch equations for a lambda-type
three-level system \cite{Fleischhauer2005,Debatin2011} in the limit of
$\Omega_{\rm FB} \ll \Omega_{\rm BB}$,
\begin{equation}
\frac{N}{N_{0}}=\exp\left(-\frac{t_{\mathrm{PA}}\Omega_{\mathrm{FB}}^{2}(4\Gamma\Delta_{2\gamma}^{2}+\Gamma_{\mathrm{eff}}(\Omega_{\mathrm{BB}}^{2}+\Gamma_{\mathrm{eff}}\Gamma))}{|\Omega_{\mathrm{BB}}^{2}+(\Gamma+2i\Delta_{\rm FB})(\Gamma_{\mathrm{eff}}+2i\Delta_{2\gamma})|^{2}}\right).\label{eq:three-level}
\end{equation}
Here, $t_{\mathrm{PA}}$ is the irradiation time of the photoassociation lasers,
$\Omega_{\mathrm{FB}}$ $(\Omega_{\mathrm{BB}})$ is the Rabi frequency on the
free-bound (bound-bound) transition, $\Delta_{2\gamma}$ is the detuning from
two-photon resonance, $\Gamma$ is the power-broadened linewidth of the
free-bound transition and $\Gamma_{\mathrm{eff}}$ is a phenomenological
constant that accounts for the decoherence of the dark state.

Figure \ref{fig:2PA spectrum}(b) shows the dark-resonance spectrum observed
when $\omega_{1}$ is resonant with the free-bound transition
$(\Delta_{\mathrm{FB}}=0)$ and $\omega_{2}$ is scanned. This complements the
EIT lineshape shown in Fig.\ \ref{fig:2PA spectrum}(a); the only difference is
which laser frequency is scanned. Off resonance with the bound-bound
transition, we observe a large loss of Cs atoms due to the production of Cs*Yb
molecules \footnote{The production of Cs*Yb molecules causes a detectable loss
of Cs atoms from the trap}. When $L_{2}$ is tuned close to resonance with the
bound-bound transition, the photon-dressed ground state and the excited state
couple to form two dressed states \cite{Fleischhauer2005}. The splitting of the
dressed states creates a dark state where $L_{1}$ is no longer resonant with
the free-bound transition. Therefore, the production of Cs*Yb molecules is
suppressed and there is a recovery in the Cs number. In the perturbative limit,
Eq.\ \ref{eq:three-level} reduces to a Lorentzian profile with a width
proportional to $\Omega_{\rm BB}^{2}$ and we therefore fit the data with a
Lorenztian lineshape. This dark-resonance technique is the simplest method for
the observation of a two-photon resonance, as with sufficient $L_2$ intensity
the feature can be significantly broadened without shifting the line center.
However, the background number of Cs atoms is sensitive to the one-photon
photoassociation loss rate and can therefore drift in response to changes in
the Yb density, the Cs density, or the photoassociation light intensity or
polarization.

Figure \ref{fig:2PA spectrum}(c) shows an alternative method for observing the
two-photon resonance using Raman spectroscopy. In this case, $\omega_{1}$ is
detuned from the free-bound transition ($\Delta_{\mathrm{FB}} = - 15 \,
\mathrm{MHz}$) and $L_{2}$ drives a stimulated Raman transition to a
vibrational level of the electronic ground state when the Raman condition is
fulfilled ($\Delta_{\mathrm{FB}} = \Delta_{\mathrm{BB}}$). This gives a narrow
lineshape. The creation of a ground-state CsYb molecule, which is dark to our
imaging, causes a decrease in the number of observed Cs atoms. The asymmetric
lineshape originates from the interference between the two paths $(E_{\rm th}
\rightarrow E_{\rm b_{2}}$ and $E_{\rm th} \rightarrow E_{\rm b_{2}}
\rightarrow E_{\rm b_{1}} \rightarrow E_{\rm b_{2}})$
\cite{Bohn1996,Portier2009} and incorporates a Fano profile \cite{Fano1961}.

We use Raman spectroscopy as the primary method for the observation of $n''=-1$
levels, as the lineshape of the two-photon feature is narrow for low powers of
$L_{1}$ and $L_{2}$. However, coupling of the ground and excited states by
$L_{1}$ and $L_{2}$ causes light shifts in both levels that are linear in laser
intensity in the perturbative limit \cite{Portier2009}. Figure \ref{fig:Laser
shift} shows the shifts $\delta_1(I_1)$ and $\delta_2(I_2)$ of the two-photon
resonance position as functions of the intensities of lasers $L_{1}$ and
$L_{2}$. We fit a straight line to the data to extract the line position at
zero intensity. As expected, the gradient of the shift with respect to
intensity is larger for the bound-bound transition, due to the larger
Franck-Condon factor (FCF) between two bound states than between a bound state
and a scattering state.

\begin{figure}
		\includegraphics[width=0.95\linewidth]{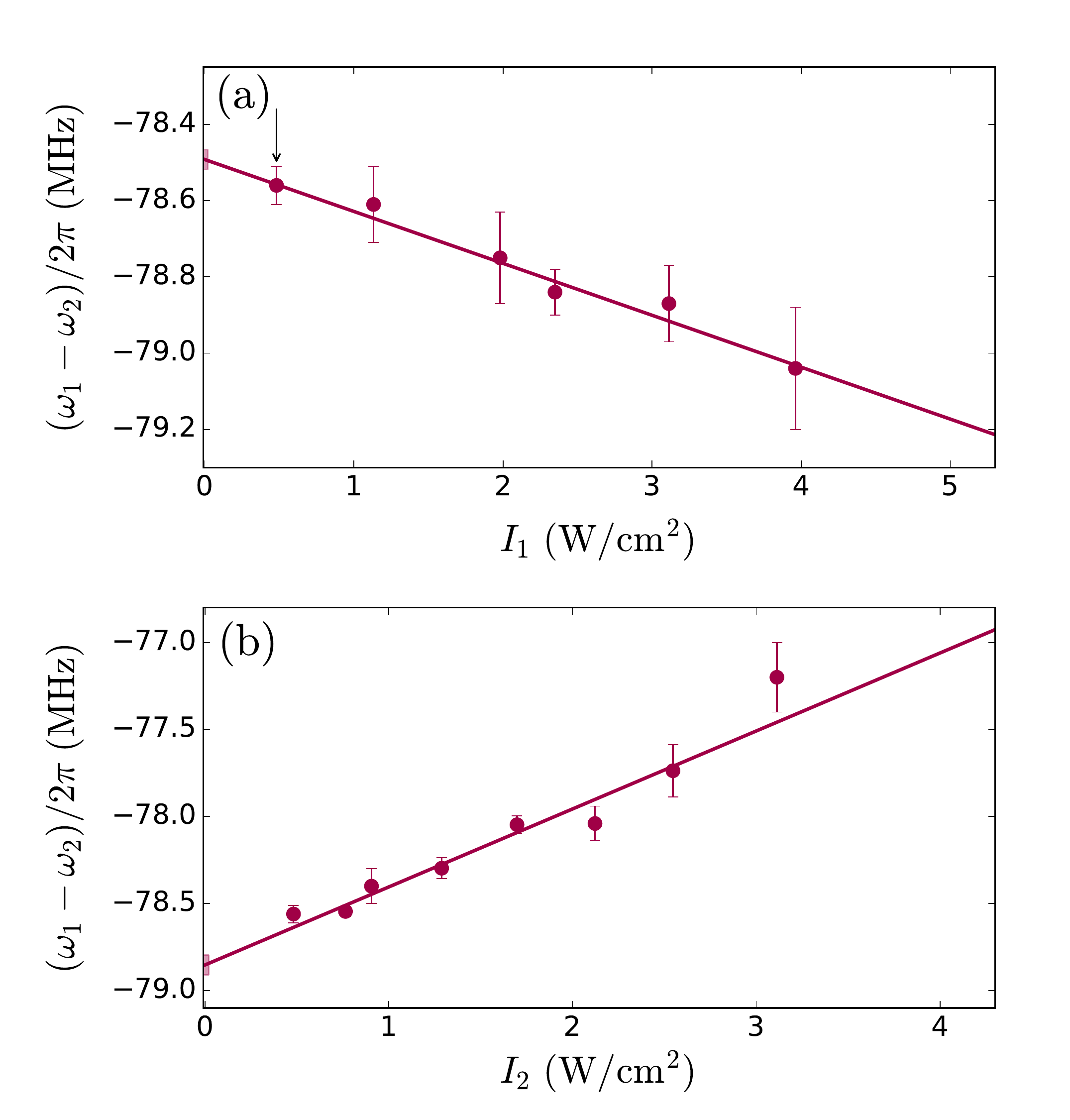}
\caption{Light shift of the Cs$^{174}$Yb $n''=-1$ Raman
line as a function of photoassociation laser intensity, using the $n'=-13$
intermediate state. (a) Measured line center frequency as a function of
intensity $I_{1}$ of laser $L_1$ driving the free-bound transition. The
intensity of laser $L_2$ for this data set was $I_{2} = 0.35 \, {\rm
W/cm^{2}}$. (b) Measured line center frequency as a function of intensity
$I_{2}$ of laser $L_2$ driving the bound-bound transition. The intensity of
laser $L_1$ for this data set was $I_{1} = 0.48 \, {\rm W/cm^{2}}$ and is
highlighted in (a). The $1 \sigma$ uncertainties in the intercepts are
represented by the shaded regions at the origins.
 \label{fig:Laser shift}}
\end{figure}

Further systematic effects that may shift the position of the Raman line are
the ac Stark shift due to the dipole trapping light, the Zeeman effect due to
the magnetic field and the finite energy of the initial atomic collision. The
trapping light may systematically shift the line position by a differential ac
Stark shift between the atomic pair and the molecular state $E_{\rm b1}$.
However, this shift is expected to be small for the weakly bound states
considered here. The effect of magnetic field on the results is small, as the
linear Zeeman shift is almost the same for the atomic state and the molecular
state. Investigation of shifts due to both magnetic field and dipole trap
intensity found no significant shift at the resolution of the measurements ($<
100 \,$kHz). The remaining systematic shift is the thermal shift, $E_{\rm th}$,
due to the energy of the initial collision between the Cs and Yb atoms. We
account for this by subtracting the mean collision energy $E_{\rm th} =
\frac{3}{2} \mu k_{\rm B}  \left(T_{\rm Yb}/m_{\rm Yb} + T_{\rm Cs}/m_{\rm Cs}
\right)$, where $\mu$ is the reduced mass. For our initial temperatures of
$T_{\rm Yb} =1 \, \mu$K and $T_{\rm Cs} =6 \, \mu$K, the correction is of order
$100 \,$kHz and is insignificant except for the measurements of the $n''=-1$
levels.

\begin{table}
  \centering

\begin{ruledtabular}
    \begin{tabular}{ccccccc}
    Yb & \multirow{2}{*}{$n'$} & \multirow{2}{*}{$n''$}    & \multicolumn{4}{c}{$E_{\rm b1}/h$ (MHz)} \\

   Isotope	&		&		& Obs	& Uncertainty	& Calc	& Obs$-$Calc	\\

\hline
    170	& -15	& -1		& 15.7	& 0.3		& 15.6	& 0.1		\\
    170	& -15	& -3		& 1576	& 2		& 1576	& 0		\\
    170	& -15	& -4		& 4259	& 2		& 4257	& 2		\\
    170	& -15	& -5		& 8988	& 2		& 8989	& 1		\\
    173	& -13	& -1		& 56.8	& 0.2		& 57.0	& 0.2		\\
    173	& -13	& -2		& 592	& 1		& 591	& 1		\\
    173	& -13	& -3		& 2166	& 1		& 2165	& 1		\\
    174	& -13	& -1		& 78.66	& 0.09	& 78.73	& 0.07	\\
    174	& -17	& -1		& 78.7	& 0.1		& 78.7	& 0.0		\\
    174	& -17	& -2		& 686.4	& 0.7		& 686.5	& 0.1		\\
    174	& -17	& -3		& 2385.5	& 0.9		& 2384.5	& 1		\\
    174	& -17	& -4		& 5749	& 1		& 5747	& 2		\\
    174	& -17	& -5		& 11358	& 1		& 11359	& 1		\\
    174	& -17	& -6		& 19803	& 1		& 19805	& 2		\\
    174	& -17	& -7		& 31672	& 2		& 31668	& 4		\\

    \end{tabular}%
\end{ruledtabular}
\caption{Observed binding energies and their uncertainties for vibrational
levels of three different isotopologs of CsYb in its electronic ground state,
together with experimental $1\sigma$ uncertainties and binding energies
calculated from the fitted interaction potential.}
  \label{table:Binding energies}%
\end{table}%

In total we observed 14 ground-state vibrational levels for the three
isotopologs Cs$^{170}$Yb, Cs$^{173}$Yb and Cs$^{174}$Yb. The binding energies
of these levels, corrected for thermal shifts and light shifts due to $L_1$ and
$L_2$, are listed in Table~\ref{table:Binding energies}. The dark-resonance
spectroscopy method scanning $\omega_{2}$ was used for measurements of the $n''
< -1$ levels. The smaller error bars for the $n''=-1$ levels result from the
narrower Raman feature and the different method of generating the small
frequency offset between the two photons. Frequency instabilities due to
beating between the sidebands of $L_{1}$ and $L_{2}$ prevented observation of
the $n''=-2$ state of Cs$^{170}$Yb. The $n''=-1$ level of Cs$^{174}$Yb was
measured with both $n' =-13$ and $n' =-17$ as intermediate states to verify
that the measurements are of the ground electronic state and not two-photon
transitions to a higher-energy electronic state. We chose to use intermediate
states with moderately large binding energies to increase the detuning of the
photoassociation light from the Cs $D_{1}$ transition; a greater feature depth
is observed for larger detuning due to the reduction of off-resonant Cs losses
\cite{Guttridge2018}.

\section{Line strengths \& Autler-Townes Spectroscopy}

The strengths of transitions between the electronically excited state and
ground state may be determined from the light shift of the Raman spectroscopy
measurements. The systematic dependencies of Raman transitions in three-level
lambda-type systems have been studied extensively
\cite{Brewer1975,Orriols1979,Lounis1992,Bohn1996,Zanon-Willette2011}. For
atomic systems it has been shown that the light shift is proportional to
$\Omega^{2}$, where $\Omega$ is the Rabi frequency associated with either
one-photon transition \cite{Brewer1975,Orriols1979}. Investigations of
molecular systems have found that the light shift of the resonance maintains
this $\Omega^{2}$ dependence even in the presence of decay out of the
three-level system \cite{Portier2009,Cohen-Tannoudji2015}. Here we determine
the line strengths for the bound-bound transitions given by $\Omega_{\rm
BB}^{2}/ I_{2}$ using light-shift measurements of the type presented in Fig.\
\ref{fig:Laser shift}(b).

\begin{figure}
		\includegraphics[width=0.95\linewidth]{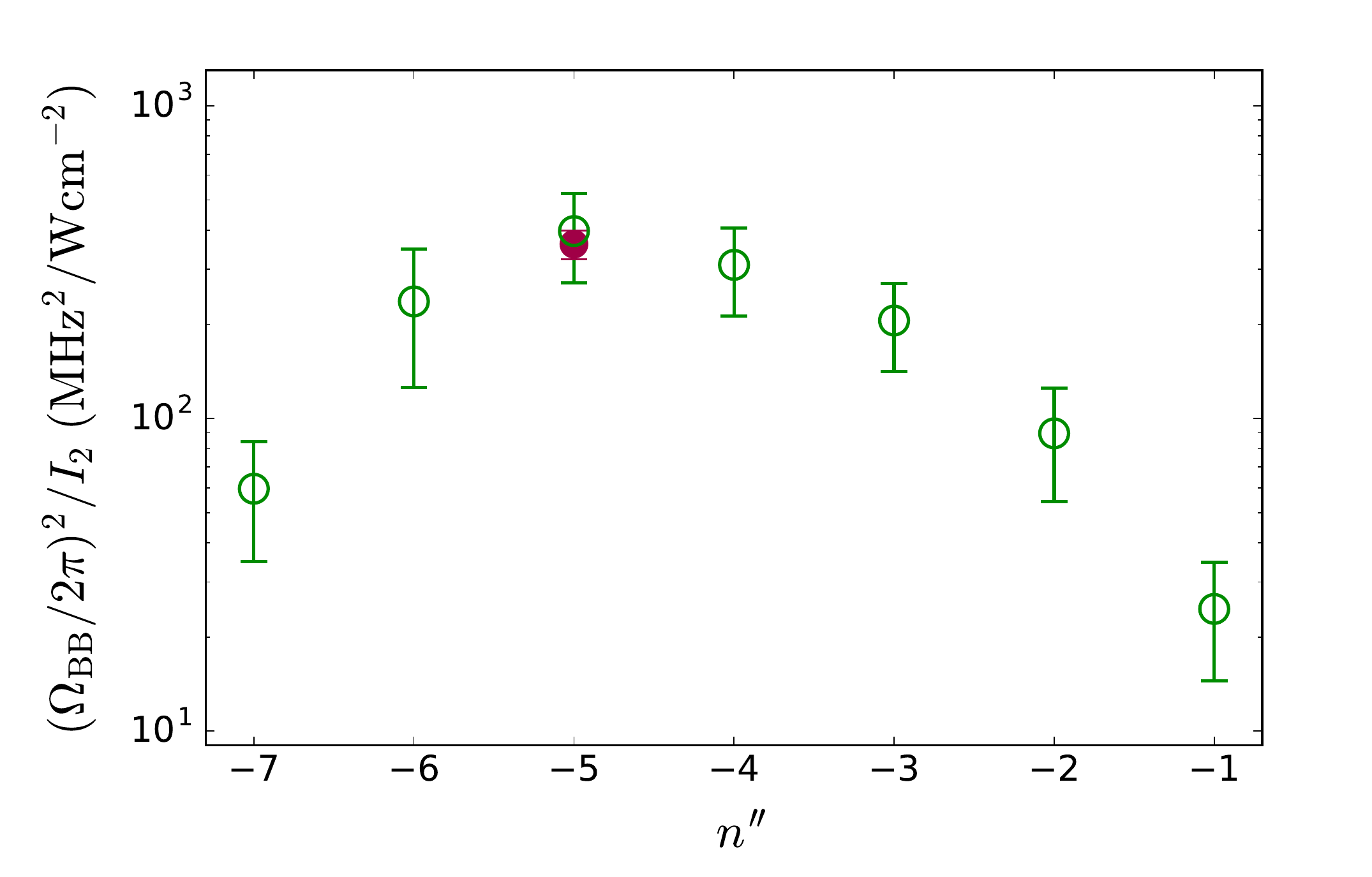}
\caption{Measured line strengths of $n'=-17 \rightarrow n''$ transitions in
Cs$^{174}$Yb. The line strength $\Omega_{\rm BB}^{2}/I_{2}$ is plotted as a
function of ground-state vibrational level $n''$. Green open circles represent
measurements of the Rabi frequencies from the line shifts of the Raman loss
features. The filled red circle represents the measurement of the $n'=-17
\rightarrow n''=-5$ transition using Autler-Townes spectroscopy.
	\label{fig:Line_strengths}}
\end{figure}

For the Raman lineshape shown in Fig.~\ref{fig:2PA spectrum}(c), the maximum
loss of Cs atoms occurs at a two-photon detuning $\omega_{1} - \omega_{2} =
E_{\rm b1}/\hbar + \delta_1(I_1) + \delta_2(I_2)$. Here $\delta_1(I_1)$ and
$\delta_2(I_2)$ are the light shifts of the transition and \cite{Portier2009}
\begin{equation}
\frac{\delta_2(I_2)}{I_2} = \left(\frac{\Omega_{\rm BB}^{2}/I_2}
{4\Delta_{\rm FB}^{2} + \Gamma^{2}}\right)\Delta_{\rm FB},\label{eq:shift}
\end{equation}
where $\Delta_{\rm FB} \simeq \Delta_{\rm BB}$ in the vicinity of the Raman
resonance \footnote{The definition of $\Omega_{\rm BB}$ in Eq.~\ref{eq:shift}
is twice that in Ref.~\cite{Portier2009}}. It follows that the line strength
$\Omega_{\rm BB}^{2}/ I_{2}$ may be obtained from the gradient of resonance
position with respect to intensity $I_{2}$ using Eq.~\ref{eq:shift}. The
results for the measured line strengths of $n'=-17 \rightarrow n''$ transitions
in Cs$^{174}$Yb are presented in Fig.\ \ref{fig:Line_strengths} as green open
circles.

\begin{figure}
		\includegraphics[width=0.95\linewidth]{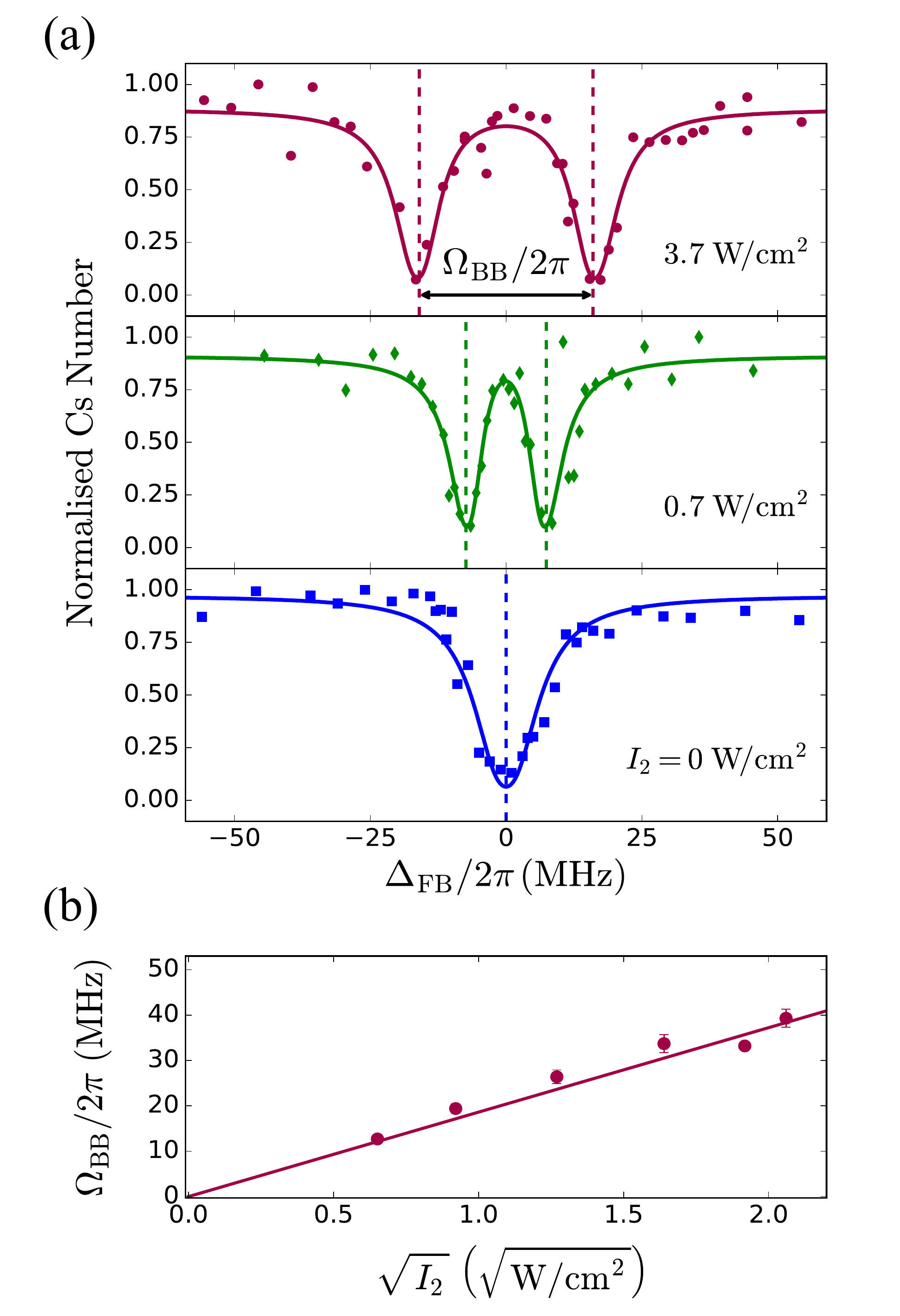}
\caption{Autler-Townes spectroscopy (ATS) of the $n'=-17 \rightarrow n''=-5$
transition in Cs$^{174}$Yb. (a) Normalized Cs number versus detuning,
$\Delta_{\rm FB}$, of laser $L_{1}$ from the $n'=-17$ free-bound transition.
The second laser $L_{2}$ is on resonance with the bound-bound transition,
$\Delta_{\rm BB} = 0$, and the splitting of the one-photon lineshape is
observed for varying intensities $I_2$ of laser $L_{2}$. (b) Bound-bound Rabi
frequency $\Omega_{\rm BB}$ extracted from the ATS measurements as a function
of the square root of the intensity $I_2$ of laser $L_2$ that drives the
bound-bound transition. The solid line is a linear fit with the intercept
constrained to be zero.
	\label{fig:ATS}}
\end{figure}

The line strengths of the bound-bound transitions may also be determined using
Autler-Townes spectroscopy (ATS) to measure the Rabi frequency, $\Omega_{\rm
BB}$, directly from the splitting of the two dressed states. The experimental
configuration for ATS is the same as in Fig.~\ref{fig:2PA spectrum}(a), but
instead of measuring the binding energy we measure the splitting of the dressed
states as a function of the intensity of $L_{2}$. Figure~\ref{fig:ATS} shows
the Autler-Townes spectrum of the $n'=-17 \rightarrow n''=-5$ transition in
Cs$^{174}$Yb. In the figure, $\omega_{2}$ is fixed on resonance ($\Delta_{\rm
BB} = 0$) and $\omega_{1}$ is scanned over the free-bound $n'=-17$ transition
for a number of different intensities of $L_{2}$. The Autler-Townes splitting
of the one-photon line is clearly visible as the intensity of the bound-bound
laser is increased. The Rabi frequency $\Omega_{\rm BB}$ is extracted by
fitting Eq.\ \ref{eq:three-level} to the data, and is approximately the
splitting of the two peaks as labeled in the figure. The quantity of interest,
$\Omega_{\rm BB} / \sqrt{I_{2}}$, is then extracted from a linear fit as shown
in Fig.~\ref{fig:ATS}(b). We find that, for the $n'=-17 \rightarrow n''=-5$
transition, $\Omega_{\rm BB} / \sqrt{I_{2}} = 2 \pi \times 19(1)\,
\mathrm{MHz}{/} \sqrt{\mathrm{W\,cm^{-2}}}$. We include this measurement in
Fig.~\ref{fig:Line_strengths} as the red closed circle. We did not measure all
the transitions using ATS due to the $\sim30$\,s load-detection cycle
associated with conducting the measurements. Nevertheless, the excellent
agreement between the two measurements of the line strength for the $n'=-17
\rightarrow n''=-5$ transition confirms the validity of using the light-shift
measurements.

The FCFs that determine the line strengths are dominated by the region around
the outermost lobe of the wavefunction for $n'=-17$. This is far inside the
outer turning points of the near-threshold levels of the ground electronic
state. In this region, the wave functions of the different near-threshold
levels in the electronic ground state are almost in phase with one another, but
with amplitudes proportional to $E_{\rm b1}^{1/3}$ \cite{LeRoy1970} and line
strengths proportional to $E_{\rm b1}^{2/3}$. However, the wave functions start
to change phase as $E_{\rm b1}$ increases; eventually the phase difference
between the wave functions in the two electronic states overcomes the amplitude
factor and the FCF starts to decrease. Figure 4 shows that the peak line
strength occurs around $n''=-5$ in the present case.

Ref.\ \cite{Guttridge2018} fitted the one-photon photoassociation spectra to a
near-dissociation expansion. However, the quantities $C_6$ and $C_8$ resulting
from this are {\em effective} dispersion coefficients that incorporate
higher-order effects. They are not sufficient to determine the outer turning
point accurately at the energy of the $n'=-17$ level, which is bound by 286
GHz. Calculating FCFs will require a more complete model of the excited-state
potential, which is beyond the scope of this paper.

\section{Determination of the Interaction Potential}

The spacings between near-threshold bound states are largely determined by the
long-range potential
\begin{equation}
%V(R) \stackrel{R\to\infty}{\sim} \sum_{n=6,8,10,\dots} -C_nR^-n,
V(R) \sim \sum_{n=6,8,10,\dots} -C_nR^{-n} \quad (\hbox{as}\ R\to\infty),
\label{eq:dispersion}
\end{equation}
where $C_n$ are dispersion coefficients. However, at least one additional
parameter is needed to specify the actual positions of the levels. To the
extent that the long-range potential is described by Eq.\ \ref{eq:dispersion},
only one such parameter is needed. This parameter may be thought of as the
binding energy of the least-bound state, the scattering length, or the
non-integer vibrational quantum number at dissociation. Physically, it is
determined by the potential at short range, and is sometimes described as the
``volume" of the potential well, as quantified by the WKB phase integral at
dissociation
\begin{equation}
\Phi = \int^{\infty}_{R_{\rm in }} \sqrt{\frac{-2 \mu}{\hbar^2}V(R)} \, dR,
\end{equation}
where $R_{\rm in }$ is the inner turning point. For a single isotopolog,
potentials with the same fractional part of $\Phi/\pi$ have the same
near-threshold bound states (and the same scattering length), even if they have
a different number of vibrational levels $N_{\rm vib}$. The Born-Oppenheimer
potential $V(R)$ is independent of reduced mass $\mu$, but the dependence of
$\Phi$ on $\mu$ means that potentials with different $N_{\rm vib}$ for one
isotopolog imply different values of the fractional part of $\Phi/\pi$, and
hence different level positions, for other isotopologs. Comparing measurements
for different isotopologs can thus establish the number of vibrational levels
supported by the potential.

Calculations of Feshbach resonance widths \cite{Zuchowski2010,Brue2013} require
a complete interaction potential $V(R)$, rather than just the long-range form,
Eq.\ \ref{eq:dispersion}. To obtain such a potential, we base the short-range
part on electronic structure calculations. Interaction potentials for the
$^2\Sigma$ ground state of CsYb have been calculated at various levels of
electronic structure theory \cite{Meyer2009, Brue2013, Shao2017,
Meniailava2017}. The potential is dominated by dispersion interactions, with
little chemical bonding, due to the large difference in ionisation energies for
Cs and Yb \footnote{3.9 eV for Cs and 6.3 eV for Yb.}. We therefore choose to
base our short-range potential on that of Brue and Hutson \cite{Brue2013}, as
the coupled-cluster methods and basis sets they used are likely to give a good
description of the dispersion interactions.

The potential of Ref.\ \cite{Brue2013} has a well depth of $hc\times 620$
cm$^{-1}$ and supports 69 vibrational levels. In order to adjust this potential
to fit our measured binding energies, we first represent it in an analytic
form,
\begin{equation}
V(R)=Ae^{-\beta R} - \sum_{n=6,8,10} D_n(\beta R) C_n R^{-n}. \label{eq:potential}
\end{equation}
Here, $A$ and $\beta$ control the magnitude and range of the short-range repulsive wall of the potential and
\begin{equation}
D_n(\beta R)=1-e^{-\beta R}\sum_{m=0}^n\frac{(\beta R)^m}{m!}
\end{equation}
is a Tang-Toennies damping function \cite{Tang1984}. To reduce the number of
free parameters, we use $C_{10}=(49/40)C_8^2/C_6$ as recommended by Thakkar and
Smith \cite{Thakkar1974}. We fit the parameters $A$, $\beta$, $C_6$ and $C_8$
to the interaction energies from the electronic structure calculations of Ref.\
\cite{Brue2013}. The functional form accurately represents the ab initio
points, and the fit is not significantly improved by including an attractive
exponential term; this confirms that there is little chemical bonding. The
value of $C_6$ obtained in this way is 3800 $E_\textrm{h}a_0^6$, which is about
13\% larger than the value of 3370 $E_\textrm{h}a_0^6$ obtained in Ref.\
\cite{Brue2013} using Tang's combination rule \cite{Tang1969}. Here, $a_0$ is
the Bohr radius and $E_\textrm{h}$ is the Hartree energy. This confirms that
the electronic structure calculations of Ref.\ \cite{Brue2013} are adequate to
give a qualitative (but not quantitative) description of the dispersion
effects.

To fit the potential to the measured binding energies, we fit the dispersion
coefficients $C_6$ and $C_8$, and vary $A$ to adjust the volume of the
potential and thus the number of vibrational levels. We fix $\beta=0.83\,
a_0^{-1}$ to the value obtained from fitting to the electronic structure
calculations. These choices allow us to fit the aspects of the potential that
are well determined by our measurements, using a small number of parameters,
while maintaining a physically reasonable form for the entire potential.

\begin{table}
  \centering

\caption{Fitted parameters and statistical uncertainties ($1\sigma$) from the
least-squares fit to the binding energies. The sensitivity is as defined in
Ref.\ \cite{leRoy1998}.}
\begin{ruledtabular}
    \begin{tabular}{cccc}
   Parameter				& Value		& Uncertainty  & Sensitivity 		\\
\hline
   $A / E_\textrm{h}$			& 13.8866515	& 0.2	& $2\times 10^{-7}$	\\
   $C_6 / E_\textrm{h}a_0^6$	& 3463.2060	    & 4		& $2\times 10^{-4}$	\\
   $C_8 / E_\textrm{h}a_0^8$	& 502560.625	& 5000  & $5\times 10^{-3}$	\\
\end{tabular}%
\end{ruledtabular}
  \label{table:Fitted_params}%
\end{table}%

We calculate near-threshold bound states supported by the potential using the
{\sc bound} package \cite{Hutson:bound:1993}. The terms in the Hamiltonian that
couple different electronic and nuclear spin channels (and cause Feshbach
resonances) are very small \cite{Brue2013}. The effective potential is thus
almost identical for all spin channels. The bound molecular states are almost
unaffected by these weak couplings. The effects of the atomic hyperfine
splitting and Zeeman shifts are already accounted for in the measurement of the
binding energies. We therefore calculate bound states using single-channel
calculations, neglecting electron and nuclear spins and the effects of the
magnetic field.

We carry out separate least-squares fits to the measured binding energies for
each plausible number of vibrational levels $N_{\rm vib}$ \footnote{In
principle, the potential might support different numbers of vibrational levels
for different isotopologs, but we find that the three isotopologs for which we
have measurements have the same number of vibrational levels.}. We fit to all
three isotopologs simultaneously, using weights derived from the experimental
uncertainties. We find the best fit for $N_{\rm vib}=77$ with a reduced
chi-squared $\chi_\nu^2=1.3$. For $N_{\rm vib}=76$ and 78 we find
$\chi_\nu^2=25$ and 26 respectively. The final fitted parameters are given in
Table \ref{table:Fitted_params}, with their uncertainties and sensitivities
\cite{leRoy1998}. As this is a very strongly correlated fit, rounding the
fitted parameters to their uncertainties introduces very large errors in the
calculated levels, so the parameters are given to a number of significant
figures determined by their sensitivity \cite{leRoy1998} to allow accurate
reproduction of the binding energies. The fitted value of $C_6$ is within 3\%
of the value from Tang's combining rule \cite{Brue2013}. The ground-state
binding energies calculated from the fitted interaction potential are included
in Table \ref{table:Binding energies}.

The statistical uncertainties in the potential parameters are very small.
However, our model is somewhat restrictive, and the uncertainties in quantities
derived from the potential are dominated by model dependence. To quantify this,
we have explored a range of different models; these include using different
values of $\beta$ and adding an attractive exponential term in the fit to the
electronic structure calculations. The estimates of uncertainties due to model
dependence given below are based on the variations observed in these tests.
Further measurements of more deeply bound vibrational states would be necessary
to determine the details of the short-range potential.

\begin{table}
  \centering
\caption{Comparison of well depth and equilibrium distances of CsYb potentials.
The uncertainties for the potential of the present work are dominated by model
dependence, not statistics.}
\begin{ruledtabular}
    \begin{tabular}{ccc}
	Ref.			& $D_{\rm e}/hc$ (cm$^{-1}$)	& $R_{\rm e}\ (a_0)$	\\
\hline
	This work		& 770(30)		        & 9.25(50)	\\
	\cite{Brue2013}	& 621					& 9.72	\\
    \cite{Shao2017}	& 542					& 9.75	\\
    \cite{Meniailava2017}	& 159			& 10.89	\\
    \cite{Meyer2009}	& 182				& 10.69 \\
\end{tabular}%
\end{ruledtabular}
  \label{table:potential_comparison}%
\end{table}

\begin{figure}
		\includegraphics[width=0.95\linewidth]{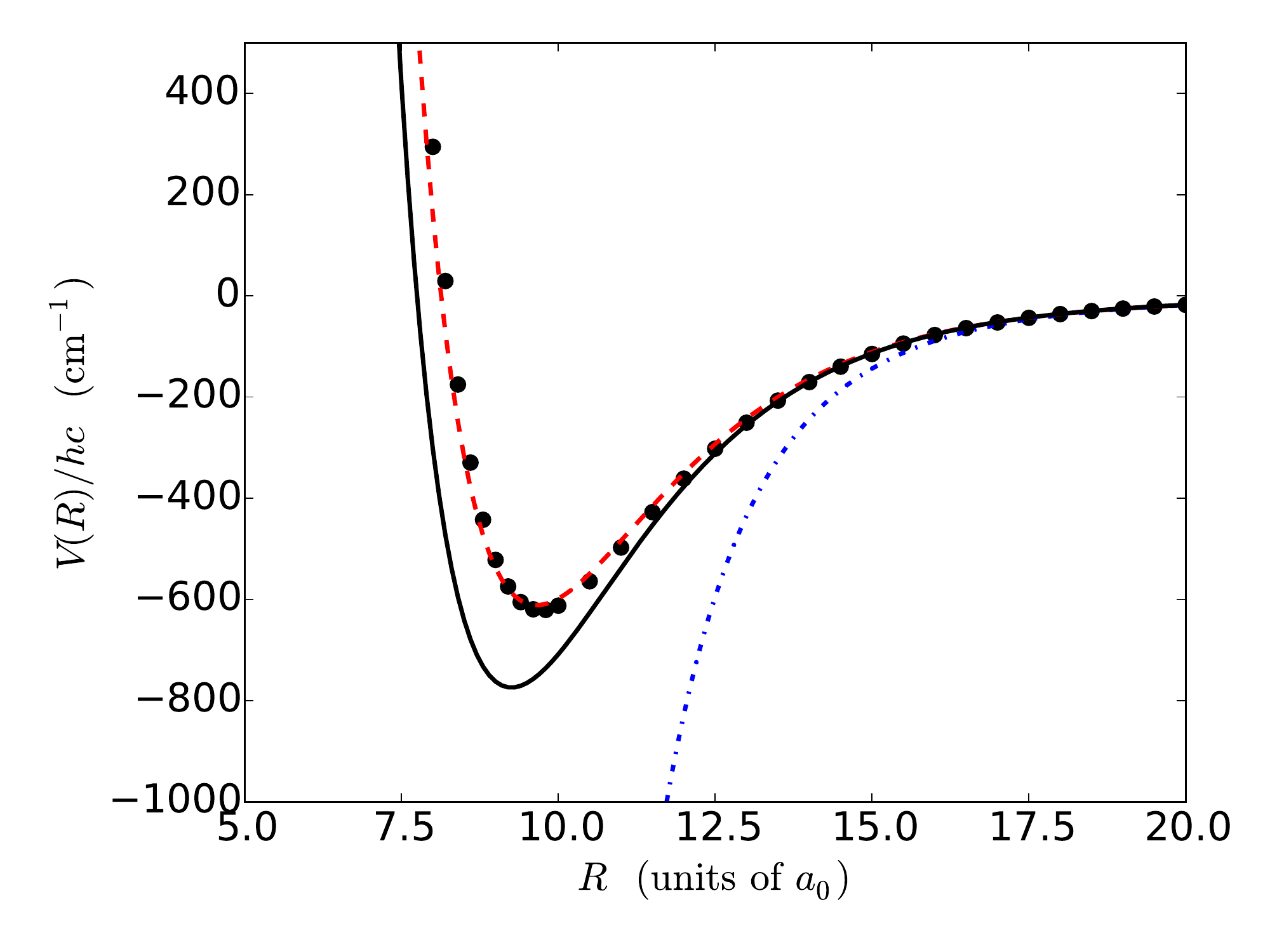}
\caption{Potential curves for the $X~^{2}\Sigma_{1/2}^{+}$ ground state of
CsYb. The dots are the electronic structure calculations of Ref.\
\cite{Brue2013}; the red dashed line is the functional form Eq.\
\eqref{eq:potential} fitted to the electronic structure calculations; the solid
black line is the final fitted potential; and the blue dash-dot line is the
pure dispersion potential, Eq.\ \eqref{eq:dispersion}, without a repulsive wall
or dispersion damping functions.
	\label{fig:potential}}
\end{figure}

\begin{figure*}
		\includegraphics[width=0.95\linewidth]{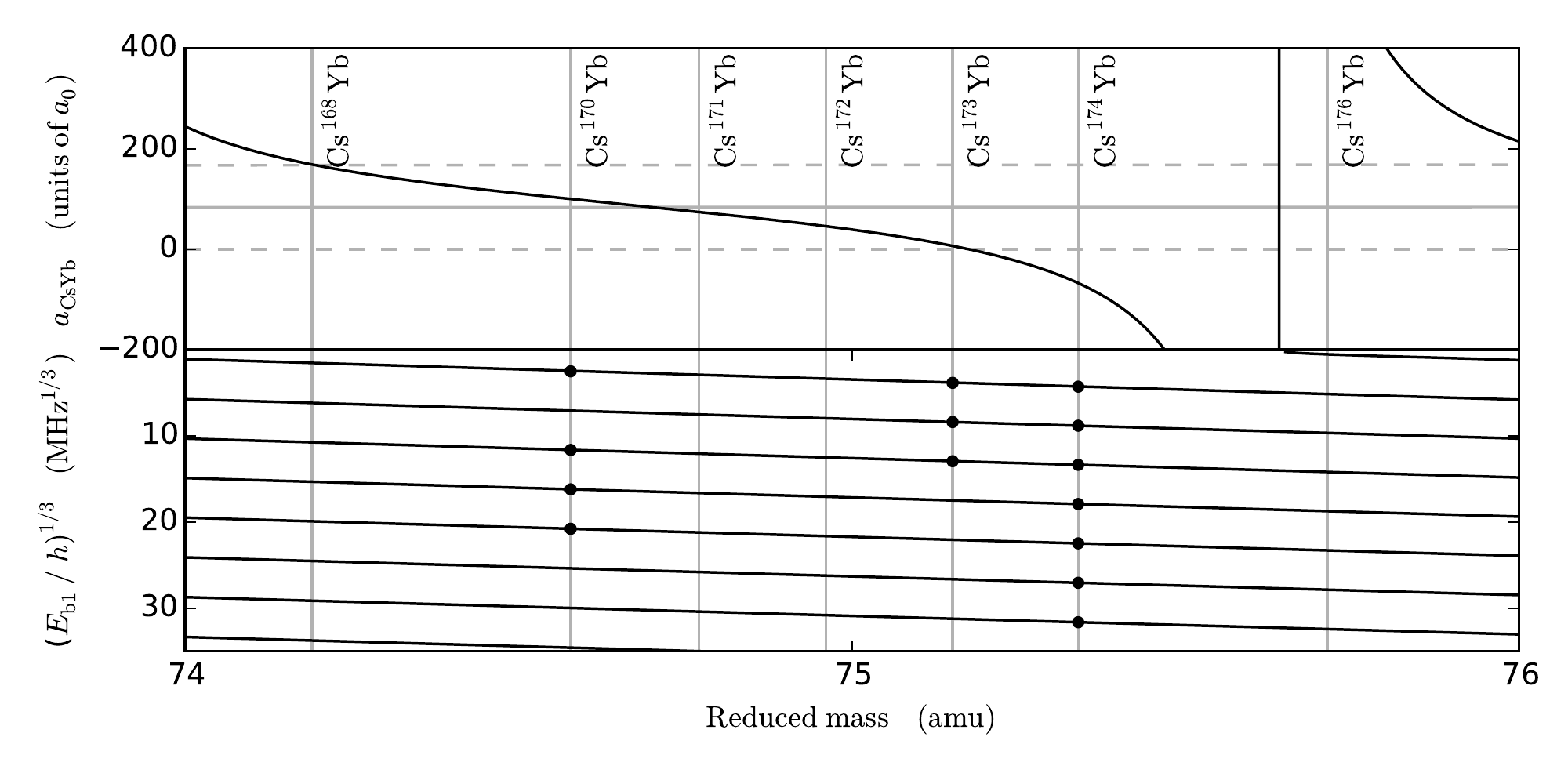}
\caption{Interspecies scattering length (upper panel) and binding energies
(lower panel) for CsYb as a function of reduced mass, calculated using the
fitted interaction potential. Points show measured levels; error bars are
smaller than the points on this scale. The vertical lines correspond to the
stable Yb isotopes. The horizontal lines on the upper figure correspond to
$a=0$, $\bar{a}$, and $2\bar{a}$.
	\label{fig:scat_ln}}
\end{figure*}

Figure \ref{fig:potential} shows the final fitted potential, along with the
unmodified potential of Brue and Hutson \cite{Brue2013}. The well depths
$D_{\rm e}$ and equilibrium distances $R_{\rm e}$ for the ground-state
potentials from Refs.\ \cite{Meyer2009}, \cite{Brue2013}, \cite{Meniailava2017}
and \cite{Shao2017} are compared with those for our fitted potential in Table
\ref{table:potential_comparison}. The minimum of our potential is deeper and at
shorter range than any of those from electronic structure calculations, though
comparable to those from Refs.\ \cite{Brue2013} and \cite{Shao2017}. There is
an inverse correlation between $D_{\rm e}$ and $R_{\rm e}$ for the different
potentials from electronic structure calculations. Refs.\ \cite{Meyer2009} and
\cite{Meniailava2017} both used large-core effective core potentials for Yb,
with only 2 active electrons; this might be responsible for their large
equilibrium distances and small well depths, which are in poor agreement with
the experimental results.

\begin{table}
  \centering
\caption{Interspecies scattering lengths calculated from the fitted interaction
potential. Both statistical uncertainties $(1\sigma)$ and estimated
uncertainties from model dependence are given.}
\begin{ruledtabular}
    \begin{tabular}{cccc}
   Mixture				& $a\ (a_0)$	& Statistical        & Model \\
        				&           	& uncertainty $(a_0)$ & dependence $(a_0)$\\
\hline
	Cs+$^{168}$Yb	& 165.98	& 0.15	& 0.4 \\
	Cs+$^{170}$Yb	& 96.24		& 0.08	& 0.2 \\
	Cs+$^{171}$Yb	& 69.99		& 0.08	& 0.3 \\
	Cs+$^{172}$Yb	& 41.03		& 0.12	& 0.5 \\
	Cs+$^{173}$Yb	& 1.0		& 0.2	& 1.0 \\
	Cs+$^{174}$Yb	& $-74.8$	& 0.5	& 3   \\
	Cs+$^{176}$Yb	& 798		& 7	    & 40  \\
\end{tabular}%
\end{ruledtabular}
  \label{table:scatlns}%
\end{table}

\section{Prediction of scattering lengths}

We have used our fitted potential to predict interspecies scattering lengths
for all isotope combinations of Cs+Yb. These are given in Table
\ref{table:scatlns}. In this case the uncertainties from statistics and model
dependence are comparable, though the latter are larger. The scattering lengths
are also shown as a function of reduced mass in Fig.\ \ref{fig:scat_ln}, along
with both observed and calculated binding energies. The cube root of the
binding energy varies almost linearly with reduced mass for an interaction
potential with $-C_6/R^6$ long-range behavior \cite{LeRoy1970}, except for a
small curvature very near dissociation due to the Gribakin-Flambaum correction
\cite{Gribakin1993} of $\pi/8$ to the WKB quantization condition at threshold.

The scattering lengths are in remarkably good agreement with our previous
estimates based on interspecies thermalization \cite{Guttridge2017}. Six of the
isotope combinations have scattering lengths between $-2\bar{a}$ and $2
\bar{a}$, where $\bar{a}$ is the mean scattering length of Gribakin and
Flambaum \cite{Gribakin1993}. The exception is Cs+$^{176}$Yb, which has a very
large scattering length due to the presence of an additional vibrational level
just below threshold. The moderate values of the scattering length for four of
the bosonic Yb isotopes should allow the production of miscible two-species
condensates \cite{Riboli2002} with Cs at the magnetic field required to
minimize the Cs three-body loss rate \cite{Weber2003}. Conversely, the large
positive scattering length for Cs+$^{176}$Yb is likely to result in an
enhancement of the widths of Feshbach resonances \cite{Brue2013}. The negative
interspecies scattering length for Cs+$^{174}$Yb opens up the intriguing
prospect of forming self-bound quantum droplets
\cite{Petrov2015,Cabrera2018,Cheiney2018}. The very small interspecies
scattering length of Cs+$^{173}$Yb indicates that the degenerate Bose-Fermi
mixture would be essentially non-interacting. In contrast, the scattering
length of $70 \, a_0$ for Cs+$^{171}$Yb is ideal for sympathetic cooling of
$^{171}$Yb to degeneracy \cite{Taie2010,Vaidya2015}, overcoming the problem of
the small intraspecies scattering length \cite{Kitagawa2008} that makes direct
evaporative cooling ineffective.

\begin{figure}
		\includegraphics[width=1\linewidth]{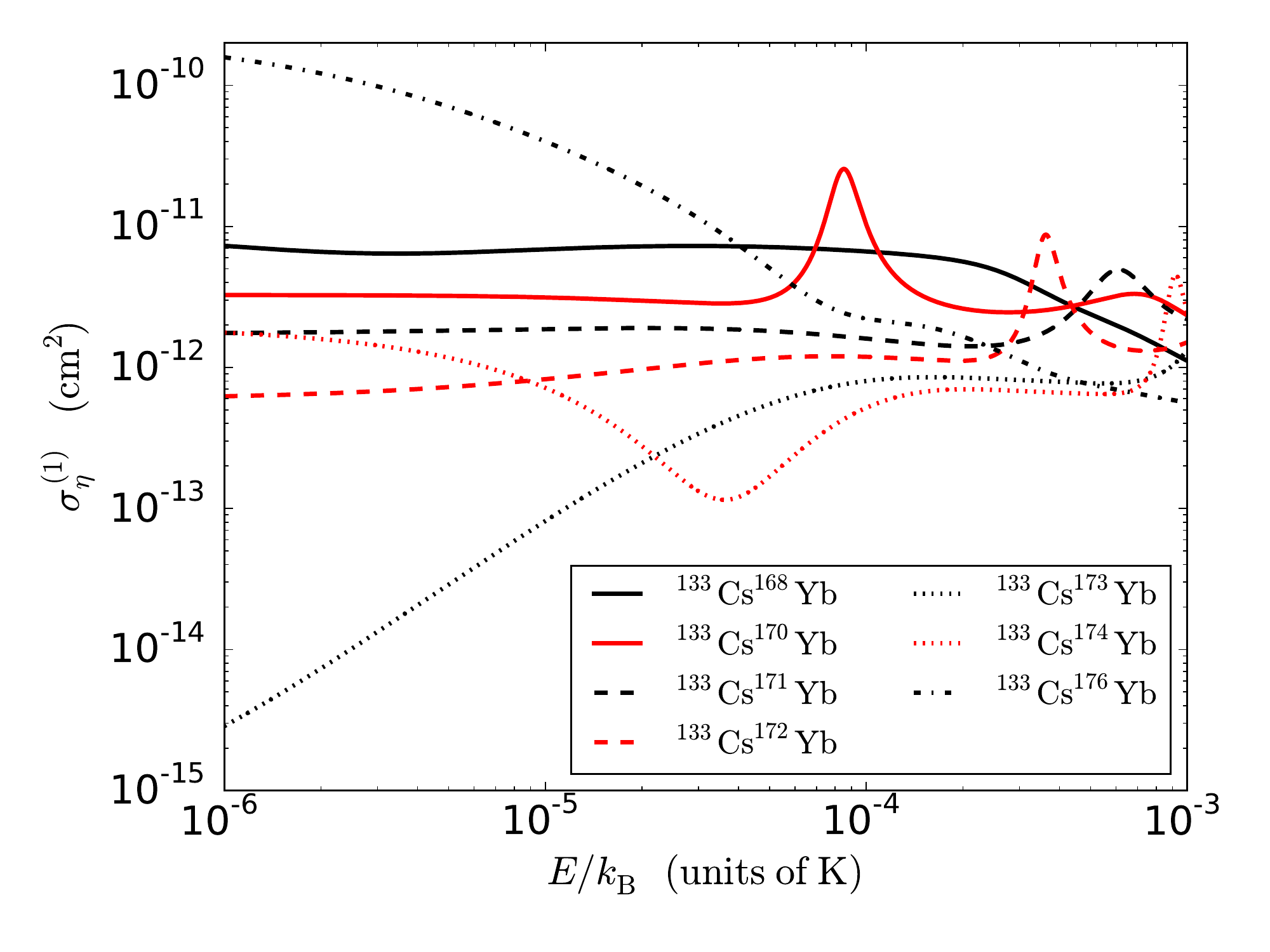}
\caption{Calculated cross sections for interspecies thermalization of Cs with
Yb, as a function of collision energy $E$.
	\label{fig:cross_sections}}
\end{figure}

Figure \ref{fig:cross_sections} shows the cross sections $\sigma_\eta^{(1)}$
that characterize interspecies thermalization \cite{Frye2014}, as a function of
collision energy, for all the isotopic combinations. These are obtained from
single-channel quantum scattering calculations on the fitted interaction
potential, using the {\sc molscat} package \cite{molscat:v14} and the
post-processor {\sc sbe} \cite{sbe}, including all relevant partial waves. The
low-energy cross sections vary across more than 4 orders of magnitude.
Cs+$^{173}$Yb has a very small cross section at low energy, due to its tiny
zero-energy scattering length, but this increases rapidly with energy due to
both effective-range effects and p-wave scattering. Cs+$^{174}$Yb has a
negative scattering length at zero energy, and exhibits a Ramsauer-Townsend
minimum near 30 $\mu$K, where the energy-dependent scattering length crosses
zero. However, the minimum is not particularly deep, because 30 $\mu$K is high
enough that the p-wave contributions are significant. Cs+$^{170}$Yb exhibits a
d-wave shape resonance around 90 $\mu$K, while Cs+$^{171}$Yb and Cs+$^{172}$Yb
exhibit f-wave shape resonances around 600 $\mu$K and 400 $\mu$K, respectively.
Cs+$^{173}$Yb and Cs+$^{174}$Yb exhibit g-wave shape resonances at even higher
energies.

\section{Conclusion}

We have used two-photon photoassociation spectroscopy to measure the binding
energies of vibrational levels of the electronic ground state of the
heteronuclear CsYb molecule. We measure the binding energy of vibrational
levels for three isotopologs of CsYb. This is sufficient to establish that the
ground state supports 77 vibrational levels. We fit a ground-state interaction
potential based on electronic structure calculations to the binding energies
for all the isotopologs together. Using our optimized potential, we calculate
values of the s-wave scattering length for all 7 isotopic combinations of
$^{133}$Cs and Yb. The results are very promising for the sympathetic cooling
of $^{171}$Yb and for the production of quantum-degenerate mixtures.

The fitted interaction potential may be used to predict positions and widths of
interspecies Feshbach resonances between a closed-shell atom and an alkali atom
\cite{Zuchowski2010,Brue2013,Barbe2017}. Magnetoassociation using these
predicted Feshbach resonances, followed by STIRAP \cite{Bergmann1998}, is a
promising route to the creation of ultracold ground-state $^{2}\Sigma$
molecules.

\begin{acknowledgments}
We acknowledge support from the UK Engineering and Physical Sciences Research
Council (grant number EP/N007085/1, EP/P008275/1 and EP/P01058X/1). The data
presented in this paper are available from
\url{http://dx.doi.org/10.15128/r1qz20ss50n}.
\end{acknowledgments}

\bibliography{TwoPhotonReferences}

\begin{thebibliography}{112}%
\makeatletter
\providecommand \@ifxundefined [1]{%
 \@ifx{#1\undefined}
}%
\providecommand \@ifnum [1]{%
 \ifnum #1\expandafter \@firstoftwo
 \else \expandafter \@secondoftwo
 \fi
}%
\providecommand \@ifx [1]{%
 \ifx #1\expandafter \@firstoftwo
 \else \expandafter \@secondoftwo
 \fi
}%
\providecommand \natexlab [1]{#1}%
\providecommand \enquote  [1]{``#1''}%
\providecommand \bibnamefont  [1]{#1}%
\providecommand \bibfnamefont [1]{#1}%
\providecommand \citenamefont [1]{#1}%
\providecommand \href@noop [0]{\@secondoftwo}%
\providecommand \href [0]{\begingroup \@sanitize@url \@href}%
\providecommand \@href[1]{\@@startlink{#1}\@@href}%
\providecommand \@@href[1]{\endgroup#1\@@endlink}%
\providecommand \@sanitize@url [0]{\catcode `\\12\catcode `\$12\catcode
  `\&12\catcode `\#12\catcode `\^12\catcode `\_12\catcode `\%12\relax}%
\providecommand \@@startlink[1]{}%
\providecommand \@@endlink[0]{}%
\providecommand \url  [0]{\begingroup\@sanitize@url \@url }%
\providecommand \@url [1]{\endgroup\@href {#1}{\urlprefix }}%
\providecommand \urlprefix  [0]{URL }%
\providecommand \Eprint [0]{\href }%
\providecommand \doibase [0]{http://dx.doi.org/}%
\providecommand \selectlanguage [0]{\@gobble}%
\providecommand \bibinfo  [0]{\@secondoftwo}%
\providecommand \bibfield  [0]{\@secondoftwo}%
\providecommand \translation [1]{[#1]}%
\providecommand \BibitemOpen [0]{}%
\providecommand \bibitemStop [0]{}%
\providecommand \bibitemNoStop [0]{.\EOS\space}%
\providecommand \EOS [0]{\spacefactor3000\relax}%
\providecommand \BibitemShut  [1]{\csname bibitem#1\endcsname}%
\let\auto@bib@innerbib\@empty
%</preamble>
\bibitem [{\citenamefont {M{\o}lmer}(1998)}]{Molmer1998}%
  \BibitemOpen
  \bibfield  {author} {\bibinfo {author} {\bibfnamefont {K.}~\bibnamefont
  {M{\o}lmer}},\ }\href {\doibase 10.1103/physrevlett.80.1804} {\bibfield
  {journal} {\bibinfo  {journal} {Phys. Rev. Lett.}\ }\textbf {\bibinfo
  {volume} {80}},\ \bibinfo {pages} {1804} (\bibinfo {year}
  {1998})}\BibitemShut {NoStop}%
\bibitem [{\citenamefont {Lewenstein}\ \emph {et~al.}(2004)\citenamefont
  {Lewenstein}, \citenamefont {Santos}, \citenamefont {Baranov},\ and\
  \citenamefont {Fehrmann}}]{Lewenstein2004a}%
  \BibitemOpen
  \bibfield  {author} {\bibinfo {author} {\bibfnamefont {M.}~\bibnamefont
  {Lewenstein}}, \bibinfo {author} {\bibfnamefont {L.}~\bibnamefont {Santos}},
  \bibinfo {author} {\bibfnamefont {M.~A.}\ \bibnamefont {Baranov}}, \ and\
  \bibinfo {author} {\bibfnamefont {H.}~\bibnamefont {Fehrmann}},\ }\href
  {\doibase 10.1103/physrevlett.92.050401} {\bibfield  {journal} {\bibinfo
  {journal} {Phys. Rev. Lett.}\ }\textbf {\bibinfo {volume} {92}},\ \bibinfo
  {pages} {050401} (\bibinfo {year} {2004})}\BibitemShut {NoStop}%
\bibitem [{\citenamefont {Zaccanti}\ \emph {et~al.}(2006)\citenamefont
  {Zaccanti}, \citenamefont {D'Errico}, \citenamefont {Ferlaino}, \citenamefont
  {Roati}, \citenamefont {Inguscio},\ and\ \citenamefont
  {Modugno}}]{Zaccanti2006a}%
  \BibitemOpen
  \bibfield  {author} {\bibinfo {author} {\bibfnamefont {M.}~\bibnamefont
  {Zaccanti}}, \bibinfo {author} {\bibfnamefont {C.}~\bibnamefont {D'Errico}},
  \bibinfo {author} {\bibfnamefont {F.}~\bibnamefont {Ferlaino}}, \bibinfo
  {author} {\bibfnamefont {G.}~\bibnamefont {Roati}}, \bibinfo {author}
  {\bibfnamefont {M.}~\bibnamefont {Inguscio}}, \ and\ \bibinfo {author}
  {\bibfnamefont {G.}~\bibnamefont {Modugno}},\ }\href {\doibase
  10.1103/physreva.74.041605} {\bibfield  {journal} {\bibinfo  {journal} {Phys.
  Rev. A}\ }\textbf {\bibinfo {volume} {74}},\ \bibinfo {pages} {041605}
  (\bibinfo {year} {2006})}\BibitemShut {NoStop}%
\bibitem [{\citenamefont {Ospelkaus}\ \emph {et~al.}(2006)\citenamefont
  {Ospelkaus}, \citenamefont {Ospelkaus}, \citenamefont {Humbert},
  \citenamefont {Sengstock},\ and\ \citenamefont {Bongs}}]{Ospelkaus2006a}%
  \BibitemOpen
  \bibfield  {author} {\bibinfo {author} {\bibfnamefont {S.}~\bibnamefont
  {Ospelkaus}}, \bibinfo {author} {\bibfnamefont {C.}~\bibnamefont
  {Ospelkaus}}, \bibinfo {author} {\bibfnamefont {L.}~\bibnamefont {Humbert}},
  \bibinfo {author} {\bibfnamefont {K.}~\bibnamefont {Sengstock}}, \ and\
  \bibinfo {author} {\bibfnamefont {K.}~\bibnamefont {Bongs}},\ }\href
  {\doibase 10.1103/physrevlett.97.120403} {\bibfield  {journal} {\bibinfo
  {journal} {Phys. Rev. Lett.}\ }\textbf {\bibinfo {volume} {97}},\ \bibinfo
  {pages} {120403} (\bibinfo {year} {2006})}\BibitemShut {NoStop}%
\bibitem [{\citenamefont {G{\"u}nter}\ \emph {et~al.}(2006)\citenamefont
  {G{\"u}nter}, \citenamefont {St{\"o}ferle}, \citenamefont {Moritz},
  \citenamefont {K{\"o}hl},\ and\ \citenamefont {Esslinger}}]{Guenter2006}%
  \BibitemOpen
  \bibfield  {author} {\bibinfo {author} {\bibfnamefont {K.}~\bibnamefont
  {G{\"u}nter}}, \bibinfo {author} {\bibfnamefont {T.}~\bibnamefont
  {St{\"o}ferle}}, \bibinfo {author} {\bibfnamefont {H.}~\bibnamefont
  {Moritz}}, \bibinfo {author} {\bibfnamefont {M.}~\bibnamefont {K{\"o}hl}}, \
  and\ \bibinfo {author} {\bibfnamefont {T.}~\bibnamefont {Esslinger}},\ }\href
  {\doibase 10.1103/physrevlett.96.180402} {\bibfield  {journal} {\bibinfo
  {journal} {Phys. Rev. Lett.}\ }\textbf {\bibinfo {volume} {96}},\ \bibinfo
  {pages} {180402} (\bibinfo {year} {2006})}\BibitemShut {NoStop}%
\bibitem [{\citenamefont {Sengupta}\ \emph {et~al.}(2007)\citenamefont
  {Sengupta}, \citenamefont {Dupuis},\ and\ \citenamefont
  {Majumdar}}]{Sengupta2007}%
  \BibitemOpen
  \bibfield  {author} {\bibinfo {author} {\bibfnamefont {K.}~\bibnamefont
  {Sengupta}}, \bibinfo {author} {\bibfnamefont {N.}~\bibnamefont {Dupuis}}, \
  and\ \bibinfo {author} {\bibfnamefont {P.}~\bibnamefont {Majumdar}},\ }\href
  {\doibase 10.1103/physreva.75.063625} {\bibfield  {journal} {\bibinfo
  {journal} {Phys. Rev. A}\ }\textbf {\bibinfo {volume} {75}},\ \bibinfo
  {pages} {063625} (\bibinfo {year} {2007})}\BibitemShut {NoStop}%
\bibitem [{\citenamefont {Marchetti}\ \emph {et~al.}(2008)\citenamefont
  {Marchetti}, \citenamefont {Mathy}, \citenamefont {Huse},\ and\ \citenamefont
  {Parish}}]{Marchetti2008}%
  \BibitemOpen
  \bibfield  {author} {\bibinfo {author} {\bibfnamefont {F.~M.}\ \bibnamefont
  {Marchetti}}, \bibinfo {author} {\bibfnamefont {C.~J.~M.}\ \bibnamefont
  {Mathy}}, \bibinfo {author} {\bibfnamefont {D.~A.}\ \bibnamefont {Huse}}, \
  and\ \bibinfo {author} {\bibfnamefont {M.~M.}\ \bibnamefont {Parish}},\
  }\href {\doibase 10.1103/physrevb.78.134517} {\bibfield  {journal} {\bibinfo
  {journal} {Phys. Rev. B}\ }\textbf {\bibinfo {volume} {78}},\ \bibinfo
  {pages} {134517} (\bibinfo {year} {2008})}\BibitemShut {NoStop}%
\bibitem [{\citenamefont {Tung}\ \emph {et~al.}(2014)\citenamefont {Tung},
  \citenamefont {Jim\'enez-Garc\'{\i}a}, \citenamefont {Johansen},
  \citenamefont {Parker},\ and\ \citenamefont {Chin}}]{Tung2014}%
  \BibitemOpen
  \bibfield  {author} {\bibinfo {author} {\bibfnamefont {S.-K.}\ \bibnamefont
  {Tung}}, \bibinfo {author} {\bibfnamefont {K.}~\bibnamefont
  {Jim\'enez-Garc\'{\i}a}}, \bibinfo {author} {\bibfnamefont {J.}~\bibnamefont
  {Johansen}}, \bibinfo {author} {\bibfnamefont {C.~V.}\ \bibnamefont
  {Parker}}, \ and\ \bibinfo {author} {\bibfnamefont {C.}~\bibnamefont
  {Chin}},\ }\href {\doibase 10.1103/PhysRevLett.113.240402} {\bibfield
  {journal} {\bibinfo  {journal} {Phys. Rev. Lett.}\ }\textbf {\bibinfo
  {volume} {113}},\ \bibinfo {pages} {240402} (\bibinfo {year}
  {2014})}\BibitemShut {NoStop}%
\bibitem [{\citenamefont {Pires}\ \emph {et~al.}(2014)\citenamefont {Pires},
  \citenamefont {Ulmanis}, \citenamefont {H\"afner}, \citenamefont {Repp},
  \citenamefont {Arias}, \citenamefont {Kuhnle},\ and\ \citenamefont
  {Weidem\"uller}}]{Pires2014}%
  \BibitemOpen
  \bibfield  {author} {\bibinfo {author} {\bibfnamefont {R.}~\bibnamefont
  {Pires}}, \bibinfo {author} {\bibfnamefont {J.}~\bibnamefont {Ulmanis}},
  \bibinfo {author} {\bibfnamefont {S.}~\bibnamefont {H\"afner}}, \bibinfo
  {author} {\bibfnamefont {M.}~\bibnamefont {Repp}}, \bibinfo {author}
  {\bibfnamefont {A.}~\bibnamefont {Arias}}, \bibinfo {author} {\bibfnamefont
  {E.~D.}\ \bibnamefont {Kuhnle}}, \ and\ \bibinfo {author} {\bibfnamefont
  {M.}~\bibnamefont {Weidem\"uller}},\ }\href {\doibase
  10.1103/PhysRevLett.112.250404} {\bibfield  {journal} {\bibinfo  {journal}
  {Phys. Rev. Lett.}\ }\textbf {\bibinfo {volume} {112}},\ \bibinfo {pages}
  {250404} (\bibinfo {year} {2014})}\BibitemShut {NoStop}%
\bibitem [{\citenamefont {Maier}\ \emph {et~al.}(2015)\citenamefont {Maier},
  \citenamefont {Eisele}, \citenamefont {Tiemann},\ and\ \citenamefont
  {Zimmermann}}]{Maier2015}%
  \BibitemOpen
  \bibfield  {author} {\bibinfo {author} {\bibfnamefont {R.~A.~W.}\
  \bibnamefont {Maier}}, \bibinfo {author} {\bibfnamefont {M.}~\bibnamefont
  {Eisele}}, \bibinfo {author} {\bibfnamefont {E.}~\bibnamefont {Tiemann}}, \
  and\ \bibinfo {author} {\bibfnamefont {C.}~\bibnamefont {Zimmermann}},\
  }\href {\doibase 10.1103/PhysRevLett.115.043201} {\bibfield  {journal}
  {\bibinfo  {journal} {Phys. Rev. Lett.}\ }\textbf {\bibinfo {volume} {115}},\
  \bibinfo {pages} {043201} (\bibinfo {year} {2015})}\BibitemShut {NoStop}%
\bibitem [{\citenamefont {Ulmanis}\ \emph {et~al.}(2016)\citenamefont
  {Ulmanis}, \citenamefont {H{\"a}fner}, \citenamefont {Pires}, \citenamefont
  {Kuhnle}, \citenamefont {Wang}, \citenamefont {Greene},\ and\ \citenamefont
  {Weidem{\"u}ller}}]{Ulmanis2016a}%
  \BibitemOpen
  \bibfield  {author} {\bibinfo {author} {\bibfnamefont {J.}~\bibnamefont
  {Ulmanis}}, \bibinfo {author} {\bibfnamefont {S.}~\bibnamefont {H{\"a}fner}},
  \bibinfo {author} {\bibfnamefont {R.}~\bibnamefont {Pires}}, \bibinfo
  {author} {\bibfnamefont {E.~D.}\ \bibnamefont {Kuhnle}}, \bibinfo {author}
  {\bibfnamefont {Y.}~\bibnamefont {Wang}}, \bibinfo {author} {\bibfnamefont
  {C.~H.}\ \bibnamefont {Greene}}, \ and\ \bibinfo {author} {\bibfnamefont
  {M.}~\bibnamefont {Weidem{\"u}ller}},\ }\href {\doibase
  10.1103/physrevlett.117.153201} {\bibfield  {journal} {\bibinfo  {journal}
  {Phys. Rev. Lett.}\ }\textbf {\bibinfo {volume} {117}},\ \bibinfo {pages}
  {153201} (\bibinfo {year} {2016})}\BibitemShut {NoStop}%
\bibitem [{\citenamefont {K{\"o}hler}\ \emph {et~al.}(2006)\citenamefont
  {K{\"o}hler}, \citenamefont {G{\'o}ral},\ and\ \citenamefont
  {Julienne}}]{Koehler2006}%
  \BibitemOpen
  \bibfield  {author} {\bibinfo {author} {\bibfnamefont {T.}~\bibnamefont
  {K{\"o}hler}}, \bibinfo {author} {\bibfnamefont {K.}~\bibnamefont
  {G{\'o}ral}}, \ and\ \bibinfo {author} {\bibfnamefont {P.~S.}\ \bibnamefont
  {Julienne}},\ }\href {\doibase 10.1103/RevModPhys.78.1311} {\bibfield
  {journal} {\bibinfo  {journal} {Rev. Mod. Phys.}\ }\textbf {\bibinfo {volume}
  {78}},\ \bibinfo {pages} {1311} (\bibinfo {year} {2006})}\BibitemShut
  {NoStop}%
\bibitem [{\citenamefont {Ni}\ \emph {et~al.}(2008)\citenamefont {Ni},
  \citenamefont {Ospelkaus}, \citenamefont {de~Miranda}, \citenamefont {Pe'er},
  \citenamefont {Neyenhuis}, \citenamefont {Zirbel}, \citenamefont
  {Kotochigova}, \citenamefont {Julienne}, \citenamefont {Jin},\ and\
  \citenamefont {Ye}}]{Ni2008}%
  \BibitemOpen
  \bibfield  {author} {\bibinfo {author} {\bibfnamefont {K.-K.}\ \bibnamefont
  {Ni}}, \bibinfo {author} {\bibfnamefont {S.}~\bibnamefont {Ospelkaus}},
  \bibinfo {author} {\bibfnamefont {M.~H.~G.}\ \bibnamefont {de~Miranda}},
  \bibinfo {author} {\bibfnamefont {A.}~\bibnamefont {Pe'er}}, \bibinfo
  {author} {\bibfnamefont {B.}~\bibnamefont {Neyenhuis}}, \bibinfo {author}
  {\bibfnamefont {J.~J.}\ \bibnamefont {Zirbel}}, \bibinfo {author}
  {\bibfnamefont {S.}~\bibnamefont {Kotochigova}}, \bibinfo {author}
  {\bibfnamefont {P.~S.}\ \bibnamefont {Julienne}}, \bibinfo {author}
  {\bibfnamefont {D.~S.}\ \bibnamefont {Jin}}, \ and\ \bibinfo {author}
  {\bibfnamefont {J.}~\bibnamefont {Ye}},\ }\href {\doibase
  10.1126/science.1163861} {\bibfield  {journal} {\bibinfo  {journal}
  {Science}\ }\textbf {\bibinfo {volume} {322}},\ \bibinfo {pages} {231}
  (\bibinfo {year} {2008})}\BibitemShut {NoStop}%
\bibitem [{\citenamefont {Lang}\ \emph {et~al.}(2008)\citenamefont {Lang},
  \citenamefont {Winkler}, \citenamefont {Strauss}, \citenamefont {Grimm},\
  and\ \citenamefont {Hecker~Denschlag}}]{Lang2008}%
  \BibitemOpen
  \bibfield  {author} {\bibinfo {author} {\bibfnamefont {F.}~\bibnamefont
  {Lang}}, \bibinfo {author} {\bibfnamefont {K.}~\bibnamefont {Winkler}},
  \bibinfo {author} {\bibfnamefont {C.}~\bibnamefont {Strauss}}, \bibinfo
  {author} {\bibfnamefont {R.}~\bibnamefont {Grimm}}, \ and\ \bibinfo {author}
  {\bibfnamefont {J.}~\bibnamefont {Hecker~Denschlag}},\ }\href {\doibase
  10.1103/PhysRevLett.101.133005} {\bibfield  {journal} {\bibinfo  {journal}
  {Phys. Rev. Lett.}\ }\textbf {\bibinfo {volume} {101}},\ \bibinfo {pages}
  {133005} (\bibinfo {year} {2008})}\BibitemShut {NoStop}%
\bibitem [{\citenamefont {Aikawa}\ \emph {et~al.}(2009)\citenamefont {Aikawa},
  \citenamefont {Akamatsu}, \citenamefont {Kobayashi}, \citenamefont {Ueda},
  \citenamefont {Kishimoto},\ and\ \citenamefont {Inouye}}]{Aikawa2009}%
  \BibitemOpen
  \bibfield  {author} {\bibinfo {author} {\bibfnamefont {K.}~\bibnamefont
  {Aikawa}}, \bibinfo {author} {\bibfnamefont {D.}~\bibnamefont {Akamatsu}},
  \bibinfo {author} {\bibfnamefont {J.}~\bibnamefont {Kobayashi}}, \bibinfo
  {author} {\bibfnamefont {M.}~\bibnamefont {Ueda}}, \bibinfo {author}
  {\bibfnamefont {T.}~\bibnamefont {Kishimoto}}, \ and\ \bibinfo {author}
  {\bibfnamefont {S.}~\bibnamefont {Inouye}},\ }\href {\doibase
  10.1088/1367-2630/11/5/055035} {\bibfield  {journal} {\bibinfo  {journal}
  {New J. Phys.}\ }\textbf {\bibinfo {volume} {11}},\ \bibinfo {pages} {055035}
  (\bibinfo {year} {2009})}\BibitemShut {NoStop}%
\bibitem [{\citenamefont {K{\"o}ppinger}\ \emph {et~al.}(2014)\citenamefont
  {K{\"o}ppinger}, \citenamefont {McCarron}, \citenamefont {Jenkin},
  \citenamefont {Molony}, \citenamefont {Cho}, \citenamefont {Cornish},
  \citenamefont {Le~Sueur}, \citenamefont {Blackley},\ and\ \citenamefont
  {Hutson}}]{Koeppinger2014}%
  \BibitemOpen
  \bibfield  {author} {\bibinfo {author} {\bibfnamefont {M.~P.}\ \bibnamefont
  {K{\"o}ppinger}}, \bibinfo {author} {\bibfnamefont {D.~J.}\ \bibnamefont
  {McCarron}}, \bibinfo {author} {\bibfnamefont {D.~L.}\ \bibnamefont
  {Jenkin}}, \bibinfo {author} {\bibfnamefont {P.~K.}\ \bibnamefont {Molony}},
  \bibinfo {author} {\bibfnamefont {H.-W.}\ \bibnamefont {Cho}}, \bibinfo
  {author} {\bibfnamefont {S.~L.}\ \bibnamefont {Cornish}}, \bibinfo {author}
  {\bibfnamefont {C.~R.}\ \bibnamefont {Le~Sueur}}, \bibinfo {author}
  {\bibfnamefont {C.~L.}\ \bibnamefont {Blackley}}, \ and\ \bibinfo {author}
  {\bibfnamefont {J.~M.}\ \bibnamefont {Hutson}},\ }\href {\doibase
  10.1103/PhysRevA.89.033604} {\bibfield  {journal} {\bibinfo  {journal} {Phys.
  Rev. A}\ }\textbf {\bibinfo {volume} {89}},\ \bibinfo {pages} {033604}
  (\bibinfo {year} {2014})}\BibitemShut {NoStop}%
\bibitem [{\citenamefont {Molony}\ \emph {et~al.}(2014)\citenamefont {Molony},
  \citenamefont {Gregory}, \citenamefont {Ji}, \citenamefont {Lu},
  \citenamefont {K\"{o}ppinger}, \citenamefont {{Le Sueur}}, \citenamefont
  {Blackley}, \citenamefont {Hutson},\ and\ \citenamefont
  {Cornish}}]{Molony2014}%
  \BibitemOpen
  \bibfield  {author} {\bibinfo {author} {\bibfnamefont {P.~K.}\ \bibnamefont
  {Molony}}, \bibinfo {author} {\bibfnamefont {P.~D.}\ \bibnamefont {Gregory}},
  \bibinfo {author} {\bibfnamefont {Z.}~\bibnamefont {Ji}}, \bibinfo {author}
  {\bibfnamefont {B.}~\bibnamefont {Lu}}, \bibinfo {author} {\bibfnamefont
  {M.~P.}\ \bibnamefont {K\"{o}ppinger}}, \bibinfo {author} {\bibfnamefont
  {C.~R.}\ \bibnamefont {{Le Sueur}}}, \bibinfo {author} {\bibfnamefont
  {C.~L.}\ \bibnamefont {Blackley}}, \bibinfo {author} {\bibfnamefont {J.~M.}\
  \bibnamefont {Hutson}}, \ and\ \bibinfo {author} {\bibfnamefont {S.~L.}\
  \bibnamefont {Cornish}},\ }\href {\doibase 10.1103/PhysRevLett.113.255301}
  {\bibfield  {journal} {\bibinfo  {journal} {Phys. Rev. Lett.}\ }\textbf
  {\bibinfo {volume} {113}},\ \bibinfo {pages} {255301} (\bibinfo {year}
  {2014})}\BibitemShut {NoStop}%
\bibitem [{\citenamefont {Molony}\ \emph {et~al.}(2016)\citenamefont {Molony},
  \citenamefont {Kumar}, \citenamefont {Gregory}, \citenamefont {Kliese},
  \citenamefont {Puppe}, \citenamefont {Le~Sueur}, \citenamefont {Aldegunde},
  \citenamefont {Hutson},\ and\ \citenamefont {Cornish}}]{Molony2016}%
  \BibitemOpen
  \bibfield  {author} {\bibinfo {author} {\bibfnamefont {P.~K.}\ \bibnamefont
  {Molony}}, \bibinfo {author} {\bibfnamefont {A.}~\bibnamefont {Kumar}},
  \bibinfo {author} {\bibfnamefont {P.~D.}\ \bibnamefont {Gregory}}, \bibinfo
  {author} {\bibfnamefont {R.}~\bibnamefont {Kliese}}, \bibinfo {author}
  {\bibfnamefont {T.}~\bibnamefont {Puppe}}, \bibinfo {author} {\bibfnamefont
  {C.~R.}\ \bibnamefont {Le~Sueur}}, \bibinfo {author} {\bibfnamefont
  {J.}~\bibnamefont {Aldegunde}}, \bibinfo {author} {\bibfnamefont {J.~M.}\
  \bibnamefont {Hutson}}, \ and\ \bibinfo {author} {\bibfnamefont {S.~L.}\
  \bibnamefont {Cornish}},\ }\href {\doibase 10.1103/PhysRevA.94.022507}
  {\bibfield  {journal} {\bibinfo  {journal} {Phys. Rev. A}\ }\textbf {\bibinfo
  {volume} {94}},\ \bibinfo {pages} {022507} (\bibinfo {year}
  {2016})}\BibitemShut {NoStop}%
\bibitem [{\citenamefont {Takekoshi}\ \emph {et~al.}(2014)\citenamefont
  {Takekoshi}, \citenamefont {Reichs\"{o}llner}, \citenamefont {Schindewolf},
  \citenamefont {Hutson}, \citenamefont {Le~Sueur}, \citenamefont {Dulieu},
  \citenamefont {Ferlaino}, \citenamefont {Grimm},\ and\ \citenamefont
  {N{\"a}gerl}}]{Takekoshi2014}%
  \BibitemOpen
  \bibfield  {author} {\bibinfo {author} {\bibfnamefont {T.}~\bibnamefont
  {Takekoshi}}, \bibinfo {author} {\bibfnamefont {L.}~\bibnamefont
  {Reichs\"{o}llner}}, \bibinfo {author} {\bibfnamefont {A.}~\bibnamefont
  {Schindewolf}}, \bibinfo {author} {\bibfnamefont {J.~M.}\ \bibnamefont
  {Hutson}}, \bibinfo {author} {\bibfnamefont {C.~R.}\ \bibnamefont
  {Le~Sueur}}, \bibinfo {author} {\bibfnamefont {O.}~\bibnamefont {Dulieu}},
  \bibinfo {author} {\bibfnamefont {F.}~\bibnamefont {Ferlaino}}, \bibinfo
  {author} {\bibfnamefont {R.}~\bibnamefont {Grimm}}, \ and\ \bibinfo {author}
  {\bibfnamefont {H.-C.}\ \bibnamefont {N{\"a}gerl}},\ }\href {\doibase
  10.1103/physrevlett.113.205301} {\bibfield  {journal} {\bibinfo  {journal}
  {Phys. Rev. Lett.}\ }\textbf {\bibinfo {volume} {113}},\ \bibinfo {pages}
  {205301} (\bibinfo {year} {2014})}\BibitemShut {NoStop}%
\bibitem [{\citenamefont {Park}\ \emph {et~al.}(2015)\citenamefont {Park},
  \citenamefont {Will},\ and\ \citenamefont {Zwierlein}}]{Park2015}%
  \BibitemOpen
  \bibfield  {author} {\bibinfo {author} {\bibfnamefont {J.~W.}\ \bibnamefont
  {Park}}, \bibinfo {author} {\bibfnamefont {S.~A.}\ \bibnamefont {Will}}, \
  and\ \bibinfo {author} {\bibfnamefont {M.~W.}\ \bibnamefont {Zwierlein}},\
  }\href {\doibase 10.1103/PhysRevLett.114.205302} {\bibfield  {journal}
  {\bibinfo  {journal} {Phys. Rev. Lett.}\ }\textbf {\bibinfo {volume} {114}},\
  \bibinfo {pages} {205302} (\bibinfo {year} {2015})}\BibitemShut {NoStop}%
\bibitem [{\citenamefont {Guo}\ \emph {et~al.}(2016)\citenamefont {Guo},
  \citenamefont {Zhu}, \citenamefont {Lu}, \citenamefont {Ye}, \citenamefont
  {Wang}, \citenamefont {Vexiau}, \citenamefont {Bouloufa-Maafa}, \citenamefont
  {Qu\'em\'ener}, \citenamefont {Dulieu},\ and\ \citenamefont
  {Wang}}]{Guo2016}%
  \BibitemOpen
  \bibfield  {author} {\bibinfo {author} {\bibfnamefont {M.}~\bibnamefont
  {Guo}}, \bibinfo {author} {\bibfnamefont {B.}~\bibnamefont {Zhu}}, \bibinfo
  {author} {\bibfnamefont {B.}~\bibnamefont {Lu}}, \bibinfo {author}
  {\bibfnamefont {X.}~\bibnamefont {Ye}}, \bibinfo {author} {\bibfnamefont
  {F.}~\bibnamefont {Wang}}, \bibinfo {author} {\bibfnamefont {R.}~\bibnamefont
  {Vexiau}}, \bibinfo {author} {\bibfnamefont {N.}~\bibnamefont
  {Bouloufa-Maafa}}, \bibinfo {author} {\bibfnamefont {G.}~\bibnamefont
  {Qu\'em\'ener}}, \bibinfo {author} {\bibfnamefont {O.}~\bibnamefont
  {Dulieu}}, \ and\ \bibinfo {author} {\bibfnamefont {D.}~\bibnamefont
  {Wang}},\ }\href {\doibase 10.1103/PhysRevLett.116.205303} {\bibfield
  {journal} {\bibinfo  {journal} {Phys. Rev. Lett.}\ }\textbf {\bibinfo
  {volume} {116}},\ \bibinfo {pages} {205303} (\bibinfo {year}
  {2016})}\BibitemShut {NoStop}%
\bibitem [{\citenamefont {Bohn}\ \emph {et~al.}(2017)\citenamefont {Bohn},
  \citenamefont {Rey},\ and\ \citenamefont {Ye}}]{Bohn2017}%
  \BibitemOpen
  \bibfield  {author} {\bibinfo {author} {\bibfnamefont {J.~L.}\ \bibnamefont
  {Bohn}}, \bibinfo {author} {\bibfnamefont {A.~M.}\ \bibnamefont {Rey}}, \
  and\ \bibinfo {author} {\bibfnamefont {J.}~\bibnamefont {Ye}},\ }\href
  {\doibase 10.1126/science.aam6299} {\bibfield  {journal} {\bibinfo  {journal}
  {Science}\ }\textbf {\bibinfo {volume} {357}},\ \bibinfo {pages} {1002}
  (\bibinfo {year} {2017})}\BibitemShut {NoStop}%
\bibitem [{\citenamefont {Modugno}\ \emph {et~al.}(2001)\citenamefont
  {Modugno}, \citenamefont {Ferrari}, \citenamefont {Roati}, \citenamefont
  {Brecha}, \citenamefont {Simoni},\ and\ \citenamefont
  {Inguscio}}]{Modugno2001}%
  \BibitemOpen
  \bibfield  {author} {\bibinfo {author} {\bibfnamefont {G.}~\bibnamefont
  {Modugno}}, \bibinfo {author} {\bibfnamefont {G.}~\bibnamefont {Ferrari}},
  \bibinfo {author} {\bibfnamefont {G.}~\bibnamefont {Roati}}, \bibinfo
  {author} {\bibfnamefont {R.~J.}\ \bibnamefont {Brecha}}, \bibinfo {author}
  {\bibfnamefont {A.}~\bibnamefont {Simoni}}, \ and\ \bibinfo {author}
  {\bibfnamefont {M.}~\bibnamefont {Inguscio}},\ }\href {\doibase
  10.1126/science.1066687} {\bibfield  {journal} {\bibinfo  {journal}
  {Science}\ }\textbf {\bibinfo {volume} {294}},\ \bibinfo {pages} {1320}
  (\bibinfo {year} {2001})}\BibitemShut {NoStop}%
\bibitem [{\citenamefont {Mudrich}\ \emph {et~al.}(2002)\citenamefont
  {Mudrich}, \citenamefont {Kraft}, \citenamefont {Singer}, \citenamefont
  {Grimm}, \citenamefont {Mosk},\ and\ \citenamefont
  {Weidem\"uller}}]{Mudrich2002}%
  \BibitemOpen
  \bibfield  {author} {\bibinfo {author} {\bibfnamefont {M.}~\bibnamefont
  {Mudrich}}, \bibinfo {author} {\bibfnamefont {S.}~\bibnamefont {Kraft}},
  \bibinfo {author} {\bibfnamefont {K.}~\bibnamefont {Singer}}, \bibinfo
  {author} {\bibfnamefont {R.}~\bibnamefont {Grimm}}, \bibinfo {author}
  {\bibfnamefont {A.}~\bibnamefont {Mosk}}, \ and\ \bibinfo {author}
  {\bibfnamefont {M.}~\bibnamefont {Weidem\"uller}},\ }\href {\doibase
  10.1103/PhysRevLett.88.253001} {\bibfield  {journal} {\bibinfo  {journal}
  {Phys. Rev. Lett.}\ }\textbf {\bibinfo {volume} {88}},\ \bibinfo {pages}
  {253001} (\bibinfo {year} {2002})}\BibitemShut {NoStop}%
\bibitem [{\citenamefont {Hadzibabic}\ \emph {et~al.}(2002)\citenamefont
  {Hadzibabic}, \citenamefont {Stan}, \citenamefont {Dieckmann}, \citenamefont
  {Gupta}, \citenamefont {Zwierlein}, \citenamefont {G\"orlitz},\ and\
  \citenamefont {Ketterle}}]{Hadzibabic2002}%
  \BibitemOpen
  \bibfield  {author} {\bibinfo {author} {\bibfnamefont {Z.}~\bibnamefont
  {Hadzibabic}}, \bibinfo {author} {\bibfnamefont {C.~A.}\ \bibnamefont
  {Stan}}, \bibinfo {author} {\bibfnamefont {K.}~\bibnamefont {Dieckmann}},
  \bibinfo {author} {\bibfnamefont {S.}~\bibnamefont {Gupta}}, \bibinfo
  {author} {\bibfnamefont {M.~W.}\ \bibnamefont {Zwierlein}}, \bibinfo {author}
  {\bibfnamefont {A.}~\bibnamefont {G\"orlitz}}, \ and\ \bibinfo {author}
  {\bibfnamefont {W.}~\bibnamefont {Ketterle}},\ }\href {\doibase
  10.1103/PhysRevLett.88.160401} {\bibfield  {journal} {\bibinfo  {journal}
  {Phys. Rev. Lett.}\ }\textbf {\bibinfo {volume} {88}},\ \bibinfo {pages}
  {160401} (\bibinfo {year} {2002})}\BibitemShut {NoStop}%
\bibitem [{\citenamefont {Taglieber}\ \emph {et~al.}(2008)\citenamefont
  {Taglieber}, \citenamefont {Voigt}, \citenamefont {Aoki}, \citenamefont
  {H\"ansch},\ and\ \citenamefont {Dieckmann}}]{Taglieber2008}%
  \BibitemOpen
  \bibfield  {author} {\bibinfo {author} {\bibfnamefont {M.}~\bibnamefont
  {Taglieber}}, \bibinfo {author} {\bibfnamefont {A.-C.}\ \bibnamefont
  {Voigt}}, \bibinfo {author} {\bibfnamefont {T.}~\bibnamefont {Aoki}},
  \bibinfo {author} {\bibfnamefont {T.~W.}\ \bibnamefont {H\"ansch}}, \ and\
  \bibinfo {author} {\bibfnamefont {K.}~\bibnamefont {Dieckmann}},\ }\href
  {\doibase 10.1103/PhysRevLett.100.010401} {\bibfield  {journal} {\bibinfo
  {journal} {Phys. Rev. Lett.}\ }\textbf {\bibinfo {volume} {100}},\ \bibinfo
  {pages} {010401} (\bibinfo {year} {2008})}\BibitemShut {NoStop}%
\bibitem [{\citenamefont {Spiegelhalder}\ \emph {et~al.}(2009)\citenamefont
  {Spiegelhalder}, \citenamefont {Trenkwalder}, \citenamefont {Naik},
  \citenamefont {Hendl}, \citenamefont {Schreck},\ and\ \citenamefont
  {Grimm}}]{Spiegelhalder2009}%
  \BibitemOpen
  \bibfield  {author} {\bibinfo {author} {\bibfnamefont {F.~M.}\ \bibnamefont
  {Spiegelhalder}}, \bibinfo {author} {\bibfnamefont {A.}~\bibnamefont
  {Trenkwalder}}, \bibinfo {author} {\bibfnamefont {D.}~\bibnamefont {Naik}},
  \bibinfo {author} {\bibfnamefont {G.}~\bibnamefont {Hendl}}, \bibinfo
  {author} {\bibfnamefont {F.}~\bibnamefont {Schreck}}, \ and\ \bibinfo
  {author} {\bibfnamefont {R.}~\bibnamefont {Grimm}},\ }\href {\doibase
  10.1103/PhysRevLett.103.223203} {\bibfield  {journal} {\bibinfo  {journal}
  {Phys. Rev. Lett.}\ }\textbf {\bibinfo {volume} {103}},\ \bibinfo {pages}
  {223203} (\bibinfo {year} {2009})}\BibitemShut {NoStop}%
\bibitem [{\citenamefont {Taie}\ \emph {et~al.}(2010)\citenamefont {Taie},
  \citenamefont {Takasu}, \citenamefont {Sugawa}, \citenamefont {Yamazaki},
  \citenamefont {Tsujimoto}, \citenamefont {Murakami},\ and\ \citenamefont
  {Takahashi}}]{Taie2010}%
  \BibitemOpen
  \bibfield  {author} {\bibinfo {author} {\bibfnamefont {S.}~\bibnamefont
  {Taie}}, \bibinfo {author} {\bibfnamefont {Y.}~\bibnamefont {Takasu}},
  \bibinfo {author} {\bibfnamefont {S.}~\bibnamefont {Sugawa}}, \bibinfo
  {author} {\bibfnamefont {R.}~\bibnamefont {Yamazaki}}, \bibinfo {author}
  {\bibfnamefont {T.}~\bibnamefont {Tsujimoto}}, \bibinfo {author}
  {\bibfnamefont {R.}~\bibnamefont {Murakami}}, \ and\ \bibinfo {author}
  {\bibfnamefont {Y.}~\bibnamefont {Takahashi}},\ }\href {\doibase
  10.1103/PhysRevLett.105.190401} {\bibfield  {journal} {\bibinfo  {journal}
  {Phys. Rev. Lett.}\ }\textbf {\bibinfo {volume} {105}},\ \bibinfo {pages}
  {190401} (\bibinfo {year} {2010})}\BibitemShut {NoStop}%
\bibitem [{\citenamefont {Cho}\ \emph {et~al.}(2011)\citenamefont {Cho},
  \citenamefont {McCarron}, \citenamefont {Jenkin}, \citenamefont
  {K{\"o}ppinger},\ and\ \citenamefont {Cornish}}]{Cho2011}%
  \BibitemOpen
  \bibfield  {author} {\bibinfo {author} {\bibfnamefont {H.~W.}\ \bibnamefont
  {Cho}}, \bibinfo {author} {\bibfnamefont {D.~J.}\ \bibnamefont {McCarron}},
  \bibinfo {author} {\bibfnamefont {D.~L.}\ \bibnamefont {Jenkin}}, \bibinfo
  {author} {\bibfnamefont {M.~P.}\ \bibnamefont {K{\"o}ppinger}}, \ and\
  \bibinfo {author} {\bibfnamefont {S.~L.}\ \bibnamefont {Cornish}},\ }\href
  {\doibase 10.1140/epjd/e2011-10716-1} {\bibfield  {journal} {\bibinfo
  {journal} {Eur. Phys. J. D}\ }\textbf {\bibinfo {volume} {65}},\ \bibinfo
  {pages} {125} (\bibinfo {year} {2011})}\BibitemShut {NoStop}%
\bibitem [{\citenamefont {McCarron}\ \emph {et~al.}(2011)\citenamefont
  {McCarron}, \citenamefont {Cho}, \citenamefont {Jenkin}, \citenamefont
  {K{\"o}ppinger},\ and\ \citenamefont {Cornish}}]{McCarron2011}%
  \BibitemOpen
  \bibfield  {author} {\bibinfo {author} {\bibfnamefont {D.~J.}\ \bibnamefont
  {McCarron}}, \bibinfo {author} {\bibfnamefont {H.~W.}\ \bibnamefont {Cho}},
  \bibinfo {author} {\bibfnamefont {D.~L.}\ \bibnamefont {Jenkin}}, \bibinfo
  {author} {\bibfnamefont {M.~P.}\ \bibnamefont {K{\"o}ppinger}}, \ and\
  \bibinfo {author} {\bibfnamefont {S.~L.}\ \bibnamefont {Cornish}},\ }\href
  {\doibase 10.1103/PhysRevA.84.011603} {\bibfield  {journal} {\bibinfo
  {journal} {Phys. Rev. A}\ }\textbf {\bibinfo {volume} {84}},\ \bibinfo
  {pages} {011603} (\bibinfo {year} {2011})}\BibitemShut {NoStop}%
\bibitem [{\citenamefont {Ridinger}\ \emph {et~al.}(2011)\citenamefont
  {Ridinger}, \citenamefont {Chaudhuri}, \citenamefont {Salez}, \citenamefont
  {Eismann}, \citenamefont {Fernandes}, \citenamefont {Magalh{\~a}es},
  \citenamefont {Wilkowski}, \citenamefont {Salomon},\ and\ \citenamefont
  {Chevy}}]{Ridinger2011}%
  \BibitemOpen
  \bibfield  {author} {\bibinfo {author} {\bibfnamefont {A.}~\bibnamefont
  {Ridinger}}, \bibinfo {author} {\bibfnamefont {S.}~\bibnamefont {Chaudhuri}},
  \bibinfo {author} {\bibfnamefont {T.}~\bibnamefont {Salez}}, \bibinfo
  {author} {\bibfnamefont {U.}~\bibnamefont {Eismann}}, \bibinfo {author}
  {\bibfnamefont {D.~R.}\ \bibnamefont {Fernandes}}, \bibinfo {author}
  {\bibfnamefont {K.}~\bibnamefont {Magalh{\~a}es}}, \bibinfo {author}
  {\bibfnamefont {D.}~\bibnamefont {Wilkowski}}, \bibinfo {author}
  {\bibfnamefont {C.}~\bibnamefont {Salomon}}, \ and\ \bibinfo {author}
  {\bibfnamefont {F.}~\bibnamefont {Chevy}},\ }\href {\doibase
  10.1140/epjd/e2011-20069-4} {\bibfield  {journal} {\bibinfo  {journal} {Eur.
  Phys. J. D}\ }\textbf {\bibinfo {volume} {65}},\ \bibinfo {pages} {223}
  (\bibinfo {year} {2011})}\BibitemShut {NoStop}%
\bibitem [{\citenamefont {Wacker}\ \emph {et~al.}(2015)\citenamefont {Wacker},
  \citenamefont {J\o{}rgensen}, \citenamefont {Birkmose}, \citenamefont
  {Horchani}, \citenamefont {Ertmer}, \citenamefont {Klempt}, \citenamefont
  {Winter}, \citenamefont {Sherson},\ and\ \citenamefont {Arlt}}]{Wacker2015}%
  \BibitemOpen
  \bibfield  {author} {\bibinfo {author} {\bibfnamefont {L.}~\bibnamefont
  {Wacker}}, \bibinfo {author} {\bibfnamefont {N.~B.}\ \bibnamefont
  {J\o{}rgensen}}, \bibinfo {author} {\bibfnamefont {D.}~\bibnamefont
  {Birkmose}}, \bibinfo {author} {\bibfnamefont {R.}~\bibnamefont {Horchani}},
  \bibinfo {author} {\bibfnamefont {W.}~\bibnamefont {Ertmer}}, \bibinfo
  {author} {\bibfnamefont {C.}~\bibnamefont {Klempt}}, \bibinfo {author}
  {\bibfnamefont {N.}~\bibnamefont {Winter}}, \bibinfo {author} {\bibfnamefont
  {J.}~\bibnamefont {Sherson}}, \ and\ \bibinfo {author} {\bibfnamefont
  {J.~J.}\ \bibnamefont {Arlt}},\ }\href {\doibase 10.1103/PhysRevA.92.053602}
  {\bibfield  {journal} {\bibinfo  {journal} {Phys. Rev. A}\ }\textbf {\bibinfo
  {volume} {92}},\ \bibinfo {pages} {053602} (\bibinfo {year}
  {2015})}\BibitemShut {NoStop}%
\bibitem [{\citenamefont {Gr\"{o}bner}\ \emph {et~al.}(2016)\citenamefont
  {Gr\"{o}bner}, \citenamefont {Weinmann}, \citenamefont {Meinert},
  \citenamefont {Lauber}, \citenamefont {Kirilov},\ and\ \citenamefont
  {N\"{a}gerl}}]{Grobner2016}%
  \BibitemOpen
  \bibfield  {author} {\bibinfo {author} {\bibfnamefont {M.}~\bibnamefont
  {Gr\"{o}bner}}, \bibinfo {author} {\bibfnamefont {P.}~\bibnamefont
  {Weinmann}}, \bibinfo {author} {\bibfnamefont {F.}~\bibnamefont {Meinert}},
  \bibinfo {author} {\bibfnamefont {K.}~\bibnamefont {Lauber}}, \bibinfo
  {author} {\bibfnamefont {E.}~\bibnamefont {Kirilov}}, \ and\ \bibinfo
  {author} {\bibfnamefont {H.-C.}\ \bibnamefont {N\"{a}gerl}},\ }\href
  {\doibase 10.1080/09500340.2016.1143051} {\bibfield  {journal} {\bibinfo
  {journal} {J. Mod. Opt.}\ }\textbf {\bibinfo {volume} {63}},\ \bibinfo
  {pages} {1829} (\bibinfo {year} {2016})}\BibitemShut {NoStop}%
\bibitem [{\citenamefont {Tassy}\ \emph {et~al.}(2010)\citenamefont {Tassy},
  \citenamefont {Nemitz}, \citenamefont {Baumer}, \citenamefont {H\"{o}hl},
  \citenamefont {Bat\"{a}r},\ and\ \citenamefont {G\"{o}rlitz}}]{Tassy2010}%
  \BibitemOpen
  \bibfield  {author} {\bibinfo {author} {\bibfnamefont {S.}~\bibnamefont
  {Tassy}}, \bibinfo {author} {\bibfnamefont {N.}~\bibnamefont {Nemitz}},
  \bibinfo {author} {\bibfnamefont {F.}~\bibnamefont {Baumer}}, \bibinfo
  {author} {\bibfnamefont {C.}~\bibnamefont {H\"{o}hl}}, \bibinfo {author}
  {\bibfnamefont {A.}~\bibnamefont {Bat\"{a}r}}, \ and\ \bibinfo {author}
  {\bibfnamefont {A.}~\bibnamefont {G\"{o}rlitz}},\ }\href
  {http://stacks.iop.org/0953-4075/43/i=20/a=205309} {\bibfield  {journal}
  {\bibinfo  {journal} {J. Phys. B: At., Mol. Opt. Phys.}\ }\textbf {\bibinfo
  {volume} {43}},\ \bibinfo {pages} {205309} (\bibinfo {year}
  {2010})}\BibitemShut {NoStop}%
\bibitem [{\citenamefont {Hara}\ \emph {et~al.}(2014)\citenamefont {Hara},
  \citenamefont {Konishi}, \citenamefont {Nakajima}, \citenamefont {Takasu},\
  and\ \citenamefont {Takahashi}}]{Hara2014}%
  \BibitemOpen
  \bibfield  {author} {\bibinfo {author} {\bibfnamefont {H.}~\bibnamefont
  {Hara}}, \bibinfo {author} {\bibfnamefont {H.}~\bibnamefont {Konishi}},
  \bibinfo {author} {\bibfnamefont {S.}~\bibnamefont {Nakajima}}, \bibinfo
  {author} {\bibfnamefont {Y.}~\bibnamefont {Takasu}}, \ and\ \bibinfo {author}
  {\bibfnamefont {Y.}~\bibnamefont {Takahashi}},\ }\href {\doibase
  10.7566/jpsj.83.014003} {\bibfield  {journal} {\bibinfo  {journal} {J. Phys.
  Soc. Jpn.}\ }\textbf {\bibinfo {volume} {83}},\ \bibinfo {pages} {014003}
  (\bibinfo {year} {2014})}\BibitemShut {NoStop}%
\bibitem [{\citenamefont {Pasquiou}\ \emph {et~al.}(2013)\citenamefont
  {Pasquiou}, \citenamefont {Bayerle}, \citenamefont {Tzanova}, \citenamefont
  {Stellmer}, \citenamefont {Szczepkowski}, \citenamefont {Parigger},
  \citenamefont {Grimm},\ and\ \citenamefont {Schreck}}]{Pasquiou2013}%
  \BibitemOpen
  \bibfield  {author} {\bibinfo {author} {\bibfnamefont {B.}~\bibnamefont
  {Pasquiou}}, \bibinfo {author} {\bibfnamefont {A.}~\bibnamefont {Bayerle}},
  \bibinfo {author} {\bibfnamefont {S.~M.}\ \bibnamefont {Tzanova}}, \bibinfo
  {author} {\bibfnamefont {S.}~\bibnamefont {Stellmer}}, \bibinfo {author}
  {\bibfnamefont {J.}~\bibnamefont {Szczepkowski}}, \bibinfo {author}
  {\bibfnamefont {M.}~\bibnamefont {Parigger}}, \bibinfo {author}
  {\bibfnamefont {R.}~\bibnamefont {Grimm}}, \ and\ \bibinfo {author}
  {\bibfnamefont {F.}~\bibnamefont {Schreck}},\ }\href {\doibase
  10.1103/PhysRevA.88.023601} {\bibfield  {journal} {\bibinfo  {journal} {Phys.
  Rev. A}\ }\textbf {\bibinfo {volume} {88}},\ \bibinfo {pages} {023601}
  (\bibinfo {year} {2013})}\BibitemShut {NoStop}%
\bibitem [{\citenamefont {Khramov}\ \emph {et~al.}(2014)\citenamefont
  {Khramov}, \citenamefont {Hansen}, \citenamefont {Dowd}, \citenamefont {Roy},
  \citenamefont {Makrides}, \citenamefont {Petrov}, \citenamefont
  {Kotochigova},\ and\ \citenamefont {Gupta}}]{Khramov2014}%
  \BibitemOpen
  \bibfield  {author} {\bibinfo {author} {\bibfnamefont {A.}~\bibnamefont
  {Khramov}}, \bibinfo {author} {\bibfnamefont {A.}~\bibnamefont {Hansen}},
  \bibinfo {author} {\bibfnamefont {W.}~\bibnamefont {Dowd}}, \bibinfo {author}
  {\bibfnamefont {R.~J.}\ \bibnamefont {Roy}}, \bibinfo {author} {\bibfnamefont
  {C.}~\bibnamefont {Makrides}}, \bibinfo {author} {\bibfnamefont
  {A.}~\bibnamefont {Petrov}}, \bibinfo {author} {\bibfnamefont
  {S.}~\bibnamefont {Kotochigova}}, \ and\ \bibinfo {author} {\bibfnamefont
  {S.}~\bibnamefont {Gupta}},\ }\href {\doibase 10.1103/PhysRevLett.112.033201}
  {\bibfield  {journal} {\bibinfo  {journal} {Phys. Rev. Lett.}\ }\textbf
  {\bibinfo {volume} {112}},\ \bibinfo {pages} {033201} (\bibinfo {year}
  {2014})}\BibitemShut {NoStop}%
\bibitem [{\citenamefont {Vaidya}\ \emph {et~al.}(2015)\citenamefont {Vaidya},
  \citenamefont {Tiamsuphat}, \citenamefont {Rolston},\ and\ \citenamefont
  {Porto}}]{Vaidya2015}%
  \BibitemOpen
  \bibfield  {author} {\bibinfo {author} {\bibfnamefont {V.~D.}\ \bibnamefont
  {Vaidya}}, \bibinfo {author} {\bibfnamefont {J.}~\bibnamefont {Tiamsuphat}},
  \bibinfo {author} {\bibfnamefont {S.~L.}\ \bibnamefont {Rolston}}, \ and\
  \bibinfo {author} {\bibfnamefont {J.~V.}\ \bibnamefont {Porto}},\ }\href
  {\doibase 10.1103/PhysRevA.92.043604} {\bibfield  {journal} {\bibinfo
  {journal} {Phys. Rev. A}\ }\textbf {\bibinfo {volume} {92}},\ \bibinfo
  {pages} {043604} (\bibinfo {year} {2015})}\BibitemShut {NoStop}%
\bibitem [{\citenamefont {Guttridge}\ \emph {et~al.}(2017)\citenamefont
  {Guttridge}, \citenamefont {Hopkins}, \citenamefont {Kemp}, \citenamefont
  {Frye}, \citenamefont {Hutson},\ and\ \citenamefont
  {Cornish}}]{Guttridge2017}%
  \BibitemOpen
  \bibfield  {author} {\bibinfo {author} {\bibfnamefont {A.}~\bibnamefont
  {Guttridge}}, \bibinfo {author} {\bibfnamefont {S.~A.}\ \bibnamefont
  {Hopkins}}, \bibinfo {author} {\bibfnamefont {S.~L.}\ \bibnamefont {Kemp}},
  \bibinfo {author} {\bibfnamefont {M.~D.}\ \bibnamefont {Frye}}, \bibinfo
  {author} {\bibfnamefont {J.~M.}\ \bibnamefont {Hutson}}, \ and\ \bibinfo
  {author} {\bibfnamefont {S.~L.}\ \bibnamefont {Cornish}},\ }\href {\doibase
  10.1103/PhysRevA.96.012704} {\bibfield  {journal} {\bibinfo  {journal} {Phys.
  Rev. A}\ }\textbf {\bibinfo {volume} {96}},\ \bibinfo {pages} {012704}
  (\bibinfo {year} {2017})}\BibitemShut {NoStop}%
\bibitem [{\citenamefont {Flores}\ \emph {et~al.}(2017)\citenamefont {Flores},
  \citenamefont {Mishra}, \citenamefont {Vassen},\ and\ \citenamefont
  {Knoop}}]{Flores2017}%
  \BibitemOpen
  \bibfield  {author} {\bibinfo {author} {\bibfnamefont {A.~S.}\ \bibnamefont
  {Flores}}, \bibinfo {author} {\bibfnamefont {H.~P.}\ \bibnamefont {Mishra}},
  \bibinfo {author} {\bibfnamefont {W.}~\bibnamefont {Vassen}}, \ and\ \bibinfo
  {author} {\bibfnamefont {S.}~\bibnamefont {Knoop}},\ }\href {\doibase
  10.1140/epjd/e2017-70675-y} {\bibfield  {journal} {\bibinfo  {journal} {Eur.
  Phys. J. D}\ }\textbf {\bibinfo {volume} {71}},\ \bibinfo {pages} {49}
  (\bibinfo {year} {2017})}\BibitemShut {NoStop}%
\bibitem [{\citenamefont {Witkowski}\ \emph {et~al.}(2017)\citenamefont
  {Witkowski}, \citenamefont {Nag\'{o}rny}, \citenamefont {Munoz-Rodriguez},
  \citenamefont {Ciury{\l}o}, \citenamefont {\.{Z}uchowski}, \citenamefont
  {Bilicki}, \citenamefont {Piotrowski}, \citenamefont {Morzy\'{n}ski},\ and\
  \citenamefont {Zawada}}]{Witkowski2017}%
  \BibitemOpen
  \bibfield  {author} {\bibinfo {author} {\bibfnamefont {M.}~\bibnamefont
  {Witkowski}}, \bibinfo {author} {\bibfnamefont {B.}~\bibnamefont
  {Nag\'{o}rny}}, \bibinfo {author} {\bibfnamefont {R.}~\bibnamefont
  {Munoz-Rodriguez}}, \bibinfo {author} {\bibfnamefont {R.}~\bibnamefont
  {Ciury{\l}o}}, \bibinfo {author} {\bibfnamefont {P.~S.}\ \bibnamefont
  {\.{Z}uchowski}}, \bibinfo {author} {\bibfnamefont {S.}~\bibnamefont
  {Bilicki}}, \bibinfo {author} {\bibfnamefont {M.}~\bibnamefont {Piotrowski}},
  \bibinfo {author} {\bibfnamefont {P.}~\bibnamefont {Morzy\'{n}ski}}, \ and\
  \bibinfo {author} {\bibfnamefont {M.}~\bibnamefont {Zawada}},\ }\href
  {\doibase 10.1364/OE.25.003165} {\bibfield  {journal} {\bibinfo  {journal}
  {Opt. Express}\ }\textbf {\bibinfo {volume} {25}},\ \bibinfo {pages} {3165}
  (\bibinfo {year} {2017})}\BibitemShut {NoStop}%
\bibitem [{\citenamefont {Micheli}\ \emph {et~al.}(2006)\citenamefont
  {Micheli}, \citenamefont {Brennen},\ and\ \citenamefont
  {Zoller}}]{Micheli2006}%
  \BibitemOpen
  \bibfield  {author} {\bibinfo {author} {\bibfnamefont {A.}~\bibnamefont
  {Micheli}}, \bibinfo {author} {\bibfnamefont {G.}~\bibnamefont {Brennen}}, \
  and\ \bibinfo {author} {\bibfnamefont {P.}~\bibnamefont {Zoller}},\ }\href
  {\doibase 10.1038/nphys287} {\bibfield  {journal} {\bibinfo  {journal} {Nat.
  Phys.}\ }\textbf {\bibinfo {volume} {2}},\ \bibinfo {pages} {341} (\bibinfo
  {year} {2006})}\BibitemShut {NoStop}%
\bibitem [{\citenamefont {P{\'e}rez-R{\'\i}os}\ \emph
  {et~al.}(2010)\citenamefont {P{\'e}rez-R{\'\i}os}, \citenamefont {Herrera},\
  and\ \citenamefont {Krems}}]{Perez-Rios2010}%
  \BibitemOpen
  \bibfield  {author} {\bibinfo {author} {\bibfnamefont {J.}~\bibnamefont
  {P{\'e}rez-R{\'\i}os}}, \bibinfo {author} {\bibfnamefont {F.}~\bibnamefont
  {Herrera}}, \ and\ \bibinfo {author} {\bibfnamefont {R.~V.}\ \bibnamefont
  {Krems}},\ }\href {http://stacks.iop.org/1367-2630/12/i=10/a=103007}
  {\bibfield  {journal} {\bibinfo  {journal} {New J. Phys.}\ }\textbf {\bibinfo
  {volume} {12}},\ \bibinfo {pages} {103007} (\bibinfo {year}
  {2010})}\BibitemShut {NoStop}%
\bibitem [{\citenamefont {Herrera}\ \emph {et~al.}(2014)\citenamefont
  {Herrera}, \citenamefont {Cao}, \citenamefont {Kais},\ and\ \citenamefont
  {Whaley}}]{Herrera2014}%
  \BibitemOpen
  \bibfield  {author} {\bibinfo {author} {\bibfnamefont {F.}~\bibnamefont
  {Herrera}}, \bibinfo {author} {\bibfnamefont {Y.}~\bibnamefont {Cao}},
  \bibinfo {author} {\bibfnamefont {S.}~\bibnamefont {Kais}}, \ and\ \bibinfo
  {author} {\bibfnamefont {K.~B.}\ \bibnamefont {Whaley}},\ }\href {\doibase
  10.1088/1367-2630/16/7/075001} {\bibfield  {journal} {\bibinfo  {journal}
  {New J. Phys.}\ }\textbf {\bibinfo {volume} {16}},\ \bibinfo {pages} {075001}
  (\bibinfo {year} {2014})}\BibitemShut {NoStop}%
\bibitem [{\citenamefont {Alyabyshev}\ \emph {et~al.}(2012)\citenamefont
  {Alyabyshev}, \citenamefont {Lemeshko},\ and\ \citenamefont
  {Krems}}]{Alyabyshev2012}%
  \BibitemOpen
  \bibfield  {author} {\bibinfo {author} {\bibfnamefont {S.~V.}\ \bibnamefont
  {Alyabyshev}}, \bibinfo {author} {\bibfnamefont {M.}~\bibnamefont
  {Lemeshko}}, \ and\ \bibinfo {author} {\bibfnamefont {R.~V.}\ \bibnamefont
  {Krems}},\ }\href {\doibase 10.1103/PhysRevA.86.013409} {\bibfield  {journal}
  {\bibinfo  {journal} {Phys. Rev. A}\ }\textbf {\bibinfo {volume} {86}},\
  \bibinfo {pages} {013409} (\bibinfo {year} {2012})}\BibitemShut {NoStop}%
\bibitem [{\citenamefont {Isaev}\ \emph {et~al.}(2010)\citenamefont {Isaev},
  \citenamefont {Hoekstra},\ and\ \citenamefont {Berger}}]{Isaev2010}%
  \BibitemOpen
  \bibfield  {author} {\bibinfo {author} {\bibfnamefont {T.~A.}\ \bibnamefont
  {Isaev}}, \bibinfo {author} {\bibfnamefont {S.}~\bibnamefont {Hoekstra}}, \
  and\ \bibinfo {author} {\bibfnamefont {R.}~\bibnamefont {Berger}},\ }\href
  {\doibase 10.1103/PhysRevA.82.052521} {\bibfield  {journal} {\bibinfo
  {journal} {Phys. Rev. A}\ }\textbf {\bibinfo {volume} {82}},\ \bibinfo
  {pages} {052521} (\bibinfo {year} {2010})}\BibitemShut {NoStop}%
\bibitem [{\citenamefont {Flambaum}\ and\ \citenamefont
  {Kozlov}(2007)}]{Flambaum2007}%
  \BibitemOpen
  \bibfield  {author} {\bibinfo {author} {\bibfnamefont {V.~V.}\ \bibnamefont
  {Flambaum}}\ and\ \bibinfo {author} {\bibfnamefont {M.~G.}\ \bibnamefont
  {Kozlov}},\ }\href {\doibase 10.1103/PhysRevLett.99.150801} {\bibfield
  {journal} {\bibinfo  {journal} {Phys. Rev. Lett.}\ }\textbf {\bibinfo
  {volume} {99}},\ \bibinfo {pages} {150801} (\bibinfo {year}
  {2007})}\BibitemShut {NoStop}%
\bibitem [{\citenamefont {Hudson}\ \emph {et~al.}(2011)\citenamefont {Hudson},
  \citenamefont {Kara}, \citenamefont {Smallman}, \citenamefont {Sauer},
  \citenamefont {Tarbutt},\ and\ \citenamefont {Hinds}}]{Hudson2011}%
  \BibitemOpen
  \bibfield  {author} {\bibinfo {author} {\bibfnamefont {J.~J.}\ \bibnamefont
  {Hudson}}, \bibinfo {author} {\bibfnamefont {D.~M.}\ \bibnamefont {Kara}},
  \bibinfo {author} {\bibfnamefont {I.~J.}\ \bibnamefont {Smallman}}, \bibinfo
  {author} {\bibfnamefont {B.~E.}\ \bibnamefont {Sauer}}, \bibinfo {author}
  {\bibfnamefont {M.~R.}\ \bibnamefont {Tarbutt}}, \ and\ \bibinfo {author}
  {\bibfnamefont {E.~A.}\ \bibnamefont {Hinds}},\ }\href {\doibase
  10.1038/nature10104} {\bibfield  {journal} {\bibinfo  {journal} {Nature}\
  }\textbf {\bibinfo {volume} {473}},\ \bibinfo {pages} {493} (\bibinfo {year}
  {2011})}\BibitemShut {NoStop}%
\bibitem [{\citenamefont {Abrahamsson}\ \emph {et~al.}(2007)\citenamefont
  {Abrahamsson}, \citenamefont {Tscherbul},\ and\ \citenamefont
  {Krems}}]{Abrahamsson2007}%
  \BibitemOpen
  \bibfield  {author} {\bibinfo {author} {\bibfnamefont {E.}~\bibnamefont
  {Abrahamsson}}, \bibinfo {author} {\bibfnamefont {T.~V.}\ \bibnamefont
  {Tscherbul}}, \ and\ \bibinfo {author} {\bibfnamefont {R.~V.}\ \bibnamefont
  {Krems}},\ }\href {\doibase 10.1063/1.2748770} {\bibfield  {journal}
  {\bibinfo  {journal} {J. Chem. Phys.}\ }\textbf {\bibinfo {volume} {127}},\
  \bibinfo {pages} {044302} (\bibinfo {year} {2007})}\BibitemShut {NoStop}%
\bibitem [{\citenamefont {Qu\'em\'ener}\ and\ \citenamefont
  {Bohn}(2016)}]{Quemener2016}%
  \BibitemOpen
  \bibfield  {author} {\bibinfo {author} {\bibfnamefont {G.}~\bibnamefont
  {Qu\'em\'ener}}\ and\ \bibinfo {author} {\bibfnamefont {J.~L.}\ \bibnamefont
  {Bohn}},\ }\href {\doibase 10.1103/PhysRevA.93.012704} {\bibfield  {journal}
  {\bibinfo  {journal} {Phys. Rev. A}\ }\textbf {\bibinfo {volume} {93}},\
  \bibinfo {pages} {012704} (\bibinfo {year} {2016})}\BibitemShut {NoStop}%
\bibitem [{\citenamefont {Hopkins}\ \emph {et~al.}(2016)\citenamefont
  {Hopkins}, \citenamefont {Butler}, \citenamefont {Guttridge}, \citenamefont
  {Kemp}, \citenamefont {Freytag}, \citenamefont {Hinds}, \citenamefont
  {Tarbutt},\ and\ \citenamefont {Cornish}}]{Hopkins2016}%
  \BibitemOpen
  \bibfield  {author} {\bibinfo {author} {\bibfnamefont {S.~A.}\ \bibnamefont
  {Hopkins}}, \bibinfo {author} {\bibfnamefont {K.}~\bibnamefont {Butler}},
  \bibinfo {author} {\bibfnamefont {A.}~\bibnamefont {Guttridge}}, \bibinfo
  {author} {\bibfnamefont {S.}~\bibnamefont {Kemp}}, \bibinfo {author}
  {\bibfnamefont {R.}~\bibnamefont {Freytag}}, \bibinfo {author} {\bibfnamefont
  {E.~A.}\ \bibnamefont {Hinds}}, \bibinfo {author} {\bibfnamefont {M.~R.}\
  \bibnamefont {Tarbutt}}, \ and\ \bibinfo {author} {\bibfnamefont {S.~L.}\
  \bibnamefont {Cornish}},\ }\href
  {http://scitation.aip.org/content/aip/journal/rsi/87/4/10.1063/1.4945795}
  {\bibfield  {journal} {\bibinfo  {journal} {Rev. Sci. Instrum.}\ }\textbf
  {\bibinfo {volume} {87}},\ \bibinfo {eid} {043109} (\bibinfo {year}
  {2016})}\BibitemShut {NoStop}%
\bibitem [{\citenamefont {Kemp}\ \emph {et~al.}(2016)\citenamefont {Kemp},
  \citenamefont {Butler}, \citenamefont {Freytag}, \citenamefont {Hopkins},
  \citenamefont {Hinds}, \citenamefont {Tarbutt},\ and\ \citenamefont
  {Cornish}}]{Kemp2016}%
  \BibitemOpen
  \bibfield  {author} {\bibinfo {author} {\bibfnamefont {S.~L.}\ \bibnamefont
  {Kemp}}, \bibinfo {author} {\bibfnamefont {K.~L.}\ \bibnamefont {Butler}},
  \bibinfo {author} {\bibfnamefont {R.}~\bibnamefont {Freytag}}, \bibinfo
  {author} {\bibfnamefont {S.~A.}\ \bibnamefont {Hopkins}}, \bibinfo {author}
  {\bibfnamefont {E.~A.}\ \bibnamefont {Hinds}}, \bibinfo {author}
  {\bibfnamefont {M.~R.}\ \bibnamefont {Tarbutt}}, \ and\ \bibinfo {author}
  {\bibfnamefont {S.~L.}\ \bibnamefont {Cornish}},\ }\href
  {http://scitation.aip.org/content/aip/journal/rsi/87/2/10.1063/1.4941719}
  {\bibfield  {journal} {\bibinfo  {journal} {Rev. Sci. Instrum.}\ }\textbf
  {\bibinfo {volume} {87}},\ \bibinfo {eid} {023105} (\bibinfo {year}
  {2016})}\BibitemShut {NoStop}%
\bibitem [{\citenamefont {Guttridge}\ \emph {et~al.}(2016)\citenamefont
  {Guttridge}, \citenamefont {Hopkins}, \citenamefont {Kemp}, \citenamefont
  {Boddy}, \citenamefont {Freytag}, \citenamefont {Jones}, \citenamefont
  {Tarbutt}, \citenamefont {Hinds},\ and\ \citenamefont
  {Cornish}}]{Guttridge2016}%
  \BibitemOpen
  \bibfield  {author} {\bibinfo {author} {\bibfnamefont {A.}~\bibnamefont
  {Guttridge}}, \bibinfo {author} {\bibfnamefont {S.~A.}\ \bibnamefont
  {Hopkins}}, \bibinfo {author} {\bibfnamefont {S.~L.}\ \bibnamefont {Kemp}},
  \bibinfo {author} {\bibfnamefont {D.}~\bibnamefont {Boddy}}, \bibinfo
  {author} {\bibfnamefont {R.}~\bibnamefont {Freytag}}, \bibinfo {author}
  {\bibfnamefont {M.~P.~A.}\ \bibnamefont {Jones}}, \bibinfo {author}
  {\bibfnamefont {M.~R.}\ \bibnamefont {Tarbutt}}, \bibinfo {author}
  {\bibfnamefont {E.~A.}\ \bibnamefont {Hinds}}, \ and\ \bibinfo {author}
  {\bibfnamefont {S.~L.}\ \bibnamefont {Cornish}},\ }\href {\doibase
  10.1088/0953-4075/49/14/145006} {\bibfield  {journal} {\bibinfo  {journal}
  {J. Phys. B: At., Mol. Opt. Phys.}\ }\textbf {\bibinfo {volume} {49}},\
  \bibinfo {pages} {145006} (\bibinfo {year} {2016})}\BibitemShut {NoStop}%
\bibitem [{\citenamefont {Chin}\ \emph {et~al.}(2010)\citenamefont {Chin},
  \citenamefont {Grimm}, \citenamefont {Julienne},\ and\ \citenamefont
  {Tiesinga}}]{Chin2010}%
  \BibitemOpen
  \bibfield  {author} {\bibinfo {author} {\bibfnamefont {C.}~\bibnamefont
  {Chin}}, \bibinfo {author} {\bibfnamefont {R.}~\bibnamefont {Grimm}},
  \bibinfo {author} {\bibfnamefont {P.}~\bibnamefont {Julienne}}, \ and\
  \bibinfo {author} {\bibfnamefont {E.}~\bibnamefont {Tiesinga}},\ }\href
  {\doibase 10.1103/RevModPhys.82.1225} {\bibfield  {journal} {\bibinfo
  {journal} {Rev. Mod. Phys.}\ }\textbf {\bibinfo {volume} {82}},\ \bibinfo
  {pages} {1225} (\bibinfo {year} {2010})}\BibitemShut {NoStop}%
\bibitem [{\citenamefont {\ifmmode~\dot{Z}\else \.{Z}\fi{}uchowski}\ \emph
  {et~al.}(2010)\citenamefont {\ifmmode~\dot{Z}\else \.{Z}\fi{}uchowski},
  \citenamefont {Aldegunde},\ and\ \citenamefont {Hutson}}]{Zuchowski2010}%
  \BibitemOpen
  \bibfield  {author} {\bibinfo {author} {\bibfnamefont {P.~S.}\ \bibnamefont
  {\ifmmode~\dot{Z}\else \.{Z}\fi{}uchowski}}, \bibinfo {author} {\bibfnamefont
  {J.}~\bibnamefont {Aldegunde}}, \ and\ \bibinfo {author} {\bibfnamefont
  {J.~M.}\ \bibnamefont {Hutson}},\ }\href {\doibase
  10.1103/PhysRevLett.105.153201} {\bibfield  {journal} {\bibinfo  {journal}
  {Phys. Rev. Lett.}\ }\textbf {\bibinfo {volume} {105}},\ \bibinfo {pages}
  {153201} (\bibinfo {year} {2010})}\BibitemShut {NoStop}%
\bibitem [{\citenamefont {Brue}\ and\ \citenamefont {Hutson}(2012)}]{Brue2012}%
  \BibitemOpen
  \bibfield  {author} {\bibinfo {author} {\bibfnamefont {D.~A.}\ \bibnamefont
  {Brue}}\ and\ \bibinfo {author} {\bibfnamefont {J.~M.}\ \bibnamefont
  {Hutson}},\ }\href {\doibase 10.1103/PhysRevLett.108.043201} {\bibfield
  {journal} {\bibinfo  {journal} {Phys. Rev. Lett.}\ }\textbf {\bibinfo
  {volume} {108}},\ \bibinfo {pages} {043201} (\bibinfo {year}
  {2012})}\BibitemShut {NoStop}%
\bibitem [{\citenamefont {Barb\'{e}}\ \emph {et~al.}(2018)\citenamefont
  {Barb\'{e}}, \citenamefont {Ciamei}, \citenamefont {Pasquiou}, \citenamefont
  {Reichs\"{o}llner}, \citenamefont {Schreck}, \citenamefont
  {\ifmmode~\dot{Z}\else \.{Z}\fi{}uchowski},\ and\ \citenamefont
  {Hutson}}]{Barbe2017}%
  \BibitemOpen
  \bibfield  {author} {\bibinfo {author} {\bibfnamefont {V.}~\bibnamefont
  {Barb\'{e}}}, \bibinfo {author} {\bibfnamefont {A.}~\bibnamefont {Ciamei}},
  \bibinfo {author} {\bibfnamefont {B.}~\bibnamefont {Pasquiou}}, \bibinfo
  {author} {\bibfnamefont {L.}~\bibnamefont {Reichs\"{o}llner}}, \bibinfo
  {author} {\bibfnamefont {F.}~\bibnamefont {Schreck}}, \bibinfo {author}
  {\bibfnamefont {P.~S.}\ \bibnamefont {\ifmmode~\dot{Z}\else
  \.{Z}\fi{}uchowski}}, \ and\ \bibinfo {author} {\bibfnamefont {J.~M.}\
  \bibnamefont {Hutson}},\ }\href {\doibase 10.1038/s41567-018-0169-x}
  {\bibfield  {journal} {\bibinfo  {journal} {Nat. Phys.}\ } (\bibinfo {year}
  {2018}),\ 10.1038/s41567-018-0169-x}\BibitemShut {NoStop}%
\bibitem [{\citenamefont {Brue}\ and\ \citenamefont {Hutson}(2013)}]{Brue2013}%
  \BibitemOpen
  \bibfield  {author} {\bibinfo {author} {\bibfnamefont {D.~A.}\ \bibnamefont
  {Brue}}\ and\ \bibinfo {author} {\bibfnamefont {J.~M.}\ \bibnamefont
  {Hutson}},\ }\href {\doibase 10.1103/PhysRevA.87.052709} {\bibfield
  {journal} {\bibinfo  {journal} {Phys. Rev. A}\ }\textbf {\bibinfo {volume}
  {87}},\ \bibinfo {pages} {052709} (\bibinfo {year} {2013})}\BibitemShut
  {NoStop}%
\bibitem [{\citenamefont {Meniailava}\ and\ \citenamefont
  {Shundalau}(2017)}]{Meniailava2017}%
  \BibitemOpen
  \bibfield  {author} {\bibinfo {author} {\bibfnamefont {D.~N.}\ \bibnamefont
  {Meniailava}}\ and\ \bibinfo {author} {\bibfnamefont {M.~B.}\ \bibnamefont
  {Shundalau}},\ }\href {\doibase 10.1016/j.comptc.2017.03.046} {\bibfield
  {journal} {\bibinfo  {journal} {Comput. Theor. Chem.}\ }\textbf {\bibinfo
  {volume} {1111}},\ \bibinfo {pages} {20} (\bibinfo {year}
  {2017})}\BibitemShut {NoStop}%
\bibitem [{\citenamefont {Bohn}\ and\ \citenamefont
  {Julienne}(1999)}]{Bohn1999}%
  \BibitemOpen
  \bibfield  {author} {\bibinfo {author} {\bibfnamefont {J.~L.}\ \bibnamefont
  {Bohn}}\ and\ \bibinfo {author} {\bibfnamefont {P.~S.}\ \bibnamefont
  {Julienne}},\ }\href {\doibase 10.1103/PhysRevA.60.414} {\bibfield  {journal}
  {\bibinfo  {journal} {Phys. Rev. A}\ }\textbf {\bibinfo {volume} {60}},\
  \bibinfo {pages} {414} (\bibinfo {year} {1999})}\BibitemShut {NoStop}%
\bibitem [{\citenamefont {Jones}\ \emph {et~al.}(2006)\citenamefont {Jones},
  \citenamefont {Tiesinga}, \citenamefont {Lett},\ and\ \citenamefont
  {Julienne}}]{Jones2006}%
  \BibitemOpen
  \bibfield  {author} {\bibinfo {author} {\bibfnamefont {K.~M.}\ \bibnamefont
  {Jones}}, \bibinfo {author} {\bibfnamefont {E.}~\bibnamefont {Tiesinga}},
  \bibinfo {author} {\bibfnamefont {P.~D.}\ \bibnamefont {Lett}}, \ and\
  \bibinfo {author} {\bibfnamefont {P.~S.}\ \bibnamefont {Julienne}},\ }\href
  {\doibase 10.1103/revmodphys.78.483} {\bibfield  {journal} {\bibinfo
  {journal} {Rev. Mod. Phys.}\ }\textbf {\bibinfo {volume} {78}},\ \bibinfo
  {pages} {483} (\bibinfo {year} {2006})}\BibitemShut {NoStop}%
\bibitem [{\citenamefont {Abraham}\ \emph {et~al.}(1996)\citenamefont
  {Abraham}, \citenamefont {McAlexander}, \citenamefont {Gerton}, \citenamefont
  {Hulet}, \citenamefont {C\^ot\'e},\ and\ \citenamefont
  {Dalgarno}}]{Abraham1996}%
  \BibitemOpen
  \bibfield  {author} {\bibinfo {author} {\bibfnamefont {E.~R.~I.}\
  \bibnamefont {Abraham}}, \bibinfo {author} {\bibfnamefont {W.~I.}\
  \bibnamefont {McAlexander}}, \bibinfo {author} {\bibfnamefont {J.~M.}\
  \bibnamefont {Gerton}}, \bibinfo {author} {\bibfnamefont {R.~G.}\
  \bibnamefont {Hulet}}, \bibinfo {author} {\bibfnamefont {R.}~\bibnamefont
  {C\^ot\'e}}, \ and\ \bibinfo {author} {\bibfnamefont {A.}~\bibnamefont
  {Dalgarno}},\ }\href {\doibase 10.1103/PhysRevA.53.R3713} {\bibfield
  {journal} {\bibinfo  {journal} {Phys. Rev. A}\ }\textbf {\bibinfo {volume}
  {53}},\ \bibinfo {pages} {R3713} (\bibinfo {year} {1996})}\BibitemShut
  {NoStop}%
\bibitem [{\citenamefont {Tsai}\ \emph {et~al.}(1997)\citenamefont {Tsai},
  \citenamefont {Freeland}, \citenamefont {Vogels}, \citenamefont {Boesten},
  \citenamefont {Verhaar},\ and\ \citenamefont {Heinzen}}]{Tsai1997}%
  \BibitemOpen
  \bibfield  {author} {\bibinfo {author} {\bibfnamefont {C.~C.}\ \bibnamefont
  {Tsai}}, \bibinfo {author} {\bibfnamefont {R.~S.}\ \bibnamefont {Freeland}},
  \bibinfo {author} {\bibfnamefont {J.~M.}\ \bibnamefont {Vogels}}, \bibinfo
  {author} {\bibfnamefont {H.~M. J.~M.}\ \bibnamefont {Boesten}}, \bibinfo
  {author} {\bibfnamefont {B.~J.}\ \bibnamefont {Verhaar}}, \ and\ \bibinfo
  {author} {\bibfnamefont {D.~J.}\ \bibnamefont {Heinzen}},\ }\href {\doibase
  10.1103/physrevlett.79.1245} {\bibfield  {journal} {\bibinfo  {journal}
  {Phys. Rev. Lett.}\ }\textbf {\bibinfo {volume} {79}},\ \bibinfo {pages}
  {1245} (\bibinfo {year} {1997})}\BibitemShut {NoStop}%
\bibitem [{\citenamefont {van Abeelen}\ and\ \citenamefont
  {Verhaar}(1999)}]{Abeelen1999a}%
  \BibitemOpen
  \bibfield  {author} {\bibinfo {author} {\bibfnamefont {F.~A.}\ \bibnamefont
  {van Abeelen}}\ and\ \bibinfo {author} {\bibfnamefont {B.~J.}\ \bibnamefont
  {Verhaar}},\ }\href {\doibase 10.1103/physreva.59.578} {\bibfield  {journal}
  {\bibinfo  {journal} {Phys. Rev. A}\ }\textbf {\bibinfo {volume} {59}},\
  \bibinfo {pages} {578} (\bibinfo {year} {1999})}\BibitemShut {NoStop}%
\bibitem [{\citenamefont {Wang}\ \emph {et~al.}(2000)\citenamefont {Wang},
  \citenamefont {Nikolov}, \citenamefont {Ensher}, \citenamefont {Gould},
  \citenamefont {Eyler}, \citenamefont {Stwalley}, \citenamefont {Burke},
  \citenamefont {Bohn}, \citenamefont {Greene}, \citenamefont {Tiesinga},
  \citenamefont {Williams},\ and\ \citenamefont {Julienne}}]{Wang2000}%
  \BibitemOpen
  \bibfield  {author} {\bibinfo {author} {\bibfnamefont {H.}~\bibnamefont
  {Wang}}, \bibinfo {author} {\bibfnamefont {A.~N.}\ \bibnamefont {Nikolov}},
  \bibinfo {author} {\bibfnamefont {J.~R.}\ \bibnamefont {Ensher}}, \bibinfo
  {author} {\bibfnamefont {P.~L.}\ \bibnamefont {Gould}}, \bibinfo {author}
  {\bibfnamefont {E.~E.}\ \bibnamefont {Eyler}}, \bibinfo {author}
  {\bibfnamefont {W.~C.}\ \bibnamefont {Stwalley}}, \bibinfo {author}
  {\bibfnamefont {J.~P.}\ \bibnamefont {Burke}}, \bibinfo {author}
  {\bibfnamefont {J.~L.}\ \bibnamefont {Bohn}}, \bibinfo {author}
  {\bibfnamefont {C.~H.}\ \bibnamefont {Greene}}, \bibinfo {author}
  {\bibfnamefont {E.}~\bibnamefont {Tiesinga}}, \bibinfo {author}
  {\bibfnamefont {C.~J.}\ \bibnamefont {Williams}}, \ and\ \bibinfo {author}
  {\bibfnamefont {P.~S.}\ \bibnamefont {Julienne}},\ }\href {\doibase
  10.1103/physreva.62.052704} {\bibfield  {journal} {\bibinfo  {journal} {Phys.
  Rev. A}\ }\textbf {\bibinfo {volume} {62}},\ \bibinfo {pages} {052704}
  (\bibinfo {year} {2000})}\BibitemShut {NoStop}%
\bibitem [{\citenamefont {Vanhaecke}\ \emph {et~al.}(2004)\citenamefont
  {Vanhaecke}, \citenamefont {Lisdat}, \citenamefont {T'Jampens}, \citenamefont
  {Comparat}, \citenamefont {Crubellier},\ and\ \citenamefont
  {Pillet}}]{Vanhaecke2004}%
  \BibitemOpen
  \bibfield  {author} {\bibinfo {author} {\bibfnamefont {N.}~\bibnamefont
  {Vanhaecke}}, \bibinfo {author} {\bibfnamefont {C.}~\bibnamefont {Lisdat}},
  \bibinfo {author} {\bibfnamefont {B.}~\bibnamefont {T'Jampens}}, \bibinfo
  {author} {\bibfnamefont {D.}~\bibnamefont {Comparat}}, \bibinfo {author}
  {\bibfnamefont {A.}~\bibnamefont {Crubellier}}, \ and\ \bibinfo {author}
  {\bibfnamefont {P.}~\bibnamefont {Pillet}},\ }\href {\doibase
  10.1140/epjd/e2004-00001-y} {\bibfield  {journal} {\bibinfo  {journal} {Eur.
  Phys. J. D}\ }\textbf {\bibinfo {volume} {28}},\ \bibinfo {pages} {351}
  (\bibinfo {year} {2004})}\BibitemShut {NoStop}%
\bibitem [{\citenamefont {Moal}\ \emph {et~al.}(2006)\citenamefont {Moal},
  \citenamefont {Portier}, \citenamefont {Kim}, \citenamefont {Dugu\'e},
  \citenamefont {Rapol}, \citenamefont {Leduc},\ and\ \citenamefont
  {Cohen-Tannoudji}}]{Moal2006}%
  \BibitemOpen
  \bibfield  {author} {\bibinfo {author} {\bibfnamefont {S.}~\bibnamefont
  {Moal}}, \bibinfo {author} {\bibfnamefont {M.}~\bibnamefont {Portier}},
  \bibinfo {author} {\bibfnamefont {J.}~\bibnamefont {Kim}}, \bibinfo {author}
  {\bibfnamefont {J.}~\bibnamefont {Dugu\'e}}, \bibinfo {author} {\bibfnamefont
  {U.~D.}\ \bibnamefont {Rapol}}, \bibinfo {author} {\bibfnamefont
  {M.}~\bibnamefont {Leduc}}, \ and\ \bibinfo {author} {\bibfnamefont
  {C.}~\bibnamefont {Cohen-Tannoudji}},\ }\href {\doibase
  10.1103/PhysRevLett.96.023203} {\bibfield  {journal} {\bibinfo  {journal}
  {Phys. Rev. Lett.}\ }\textbf {\bibinfo {volume} {96}},\ \bibinfo {pages}
  {023203} (\bibinfo {year} {2006})}\BibitemShut {NoStop}%
\bibitem [{\citenamefont {Martinez~de Escobar}\ \emph
  {et~al.}(2008)\citenamefont {Martinez~de Escobar}, \citenamefont {Mickelson},
  \citenamefont {Pellegrini}, \citenamefont {Nagel}, \citenamefont {Traverso},
  \citenamefont {Yan}, \citenamefont {C\^ot\'e},\ and\ \citenamefont
  {Killian}}]{MartinezdeEscobar2008}%
  \BibitemOpen
  \bibfield  {author} {\bibinfo {author} {\bibfnamefont {Y.~N.}\ \bibnamefont
  {Martinez~de Escobar}}, \bibinfo {author} {\bibfnamefont {P.~G.}\
  \bibnamefont {Mickelson}}, \bibinfo {author} {\bibfnamefont {P.}~\bibnamefont
  {Pellegrini}}, \bibinfo {author} {\bibfnamefont {S.~B.}\ \bibnamefont
  {Nagel}}, \bibinfo {author} {\bibfnamefont {A.}~\bibnamefont {Traverso}},
  \bibinfo {author} {\bibfnamefont {M.}~\bibnamefont {Yan}}, \bibinfo {author}
  {\bibfnamefont {R.}~\bibnamefont {C\^ot\'e}}, \ and\ \bibinfo {author}
  {\bibfnamefont {T.~C.}\ \bibnamefont {Killian}},\ }\href {\doibase
  10.1103/PhysRevA.78.062708} {\bibfield  {journal} {\bibinfo  {journal} {Phys.
  Rev. A}\ }\textbf {\bibinfo {volume} {78}},\ \bibinfo {pages} {062708}
  (\bibinfo {year} {2008})}\BibitemShut {NoStop}%
\bibitem [{\citenamefont {Kitagawa}\ \emph {et~al.}(2008)\citenamefont
  {Kitagawa}, \citenamefont {Enomoto}, \citenamefont {Kasa}, \citenamefont
  {Takahashi}, \citenamefont {Ciury{\l}o}, \citenamefont {Naidon},\ and\
  \citenamefont {Julienne}}]{Kitagawa2008}%
  \BibitemOpen
  \bibfield  {author} {\bibinfo {author} {\bibfnamefont {M.}~\bibnamefont
  {Kitagawa}}, \bibinfo {author} {\bibfnamefont {K.}~\bibnamefont {Enomoto}},
  \bibinfo {author} {\bibfnamefont {K.}~\bibnamefont {Kasa}}, \bibinfo {author}
  {\bibfnamefont {Y.}~\bibnamefont {Takahashi}}, \bibinfo {author}
  {\bibfnamefont {R.}~\bibnamefont {Ciury{\l}o}}, \bibinfo {author}
  {\bibfnamefont {P.}~\bibnamefont {Naidon}}, \ and\ \bibinfo {author}
  {\bibfnamefont {P.~S.}\ \bibnamefont {Julienne}},\ }\href {\doibase
  10.1103/physreva.77.012719} {\bibfield  {journal} {\bibinfo  {journal} {Phys.
  Rev. A}\ }\textbf {\bibinfo {volume} {77}},\ \bibinfo {pages} {012719}
  (\bibinfo {year} {2008})}\BibitemShut {NoStop}%
\bibitem [{\citenamefont {Gunton}\ \emph {et~al.}(2013)\citenamefont {Gunton},
  \citenamefont {Semczuk}, \citenamefont {Dattani},\ and\ \citenamefont
  {Madison}}]{Gunton2013}%
  \BibitemOpen
  \bibfield  {author} {\bibinfo {author} {\bibfnamefont {W.}~\bibnamefont
  {Gunton}}, \bibinfo {author} {\bibfnamefont {M.}~\bibnamefont {Semczuk}},
  \bibinfo {author} {\bibfnamefont {N.~S.}\ \bibnamefont {Dattani}}, \ and\
  \bibinfo {author} {\bibfnamefont {K.~W.}\ \bibnamefont {Madison}},\ }\href
  {\doibase 10.1103/PhysRevA.88.062510} {\bibfield  {journal} {\bibinfo
  {journal} {Phys. Rev. A}\ }\textbf {\bibinfo {volume} {88}},\ \bibinfo
  {pages} {062510} (\bibinfo {year} {2013})}\BibitemShut {NoStop}%
\bibitem [{\citenamefont {Pachomow}\ \emph {et~al.}(2017)\citenamefont
  {Pachomow}, \citenamefont {Dahlke}, \citenamefont {Tiemann}, \citenamefont
  {Riehle},\ and\ \citenamefont {Sterr}}]{Pachomow2017}%
  \BibitemOpen
  \bibfield  {author} {\bibinfo {author} {\bibfnamefont {E.}~\bibnamefont
  {Pachomow}}, \bibinfo {author} {\bibfnamefont {V.~P.}\ \bibnamefont
  {Dahlke}}, \bibinfo {author} {\bibfnamefont {E.}~\bibnamefont {Tiemann}},
  \bibinfo {author} {\bibfnamefont {F.}~\bibnamefont {Riehle}}, \ and\ \bibinfo
  {author} {\bibfnamefont {U.}~\bibnamefont {Sterr}},\ }\href {\doibase
  10.1103/PhysRevA.95.043422} {\bibfield  {journal} {\bibinfo  {journal} {Phys.
  Rev. A}\ }\textbf {\bibinfo {volume} {95}},\ \bibinfo {pages} {043422}
  (\bibinfo {year} {2017})}\BibitemShut {NoStop}%
\bibitem [{\citenamefont {M{\"u}nchow}\ \emph {et~al.}(2011)\citenamefont
  {M{\"u}nchow}, \citenamefont {Bruni}, \citenamefont {Madalinski},\ and\
  \citenamefont {Gorlitz}}]{Munchow2011}%
  \BibitemOpen
  \bibfield  {author} {\bibinfo {author} {\bibfnamefont {F.}~\bibnamefont
  {M{\"u}nchow}}, \bibinfo {author} {\bibfnamefont {C.}~\bibnamefont {Bruni}},
  \bibinfo {author} {\bibfnamefont {M.}~\bibnamefont {Madalinski}}, \ and\
  \bibinfo {author} {\bibfnamefont {A.}~\bibnamefont {Gorlitz}},\ }\href
  {\doibase 10.1039/c1cp21219b} {\bibfield  {journal} {\bibinfo  {journal}
  {Phys. Chem. Chem. Phys.}\ }\textbf {\bibinfo {volume} {13}},\ \bibinfo
  {pages} {18734} (\bibinfo {year} {2011})}\BibitemShut {NoStop}%
\bibitem [{\citenamefont {Debatin}\ \emph {et~al.}(2011)\citenamefont
  {Debatin}, \citenamefont {Takekoshi}, \citenamefont {Rameshan}, \citenamefont
  {Reichs{\"o}llner}, \citenamefont {Ferlaino}, \citenamefont {Grimm},
  \citenamefont {Vexiau}, \citenamefont {Bouloufa}, \citenamefont {Dulieu},\
  and\ \citenamefont {N{\"a}gerl}}]{Debatin2011}%
  \BibitemOpen
  \bibfield  {author} {\bibinfo {author} {\bibfnamefont {M.}~\bibnamefont
  {Debatin}}, \bibinfo {author} {\bibfnamefont {T.}~\bibnamefont {Takekoshi}},
  \bibinfo {author} {\bibfnamefont {R.}~\bibnamefont {Rameshan}}, \bibinfo
  {author} {\bibfnamefont {L.}~\bibnamefont {Reichs{\"o}llner}}, \bibinfo
  {author} {\bibfnamefont {F.}~\bibnamefont {Ferlaino}}, \bibinfo {author}
  {\bibfnamefont {R.}~\bibnamefont {Grimm}}, \bibinfo {author} {\bibfnamefont
  {R.}~\bibnamefont {Vexiau}}, \bibinfo {author} {\bibfnamefont
  {N.}~\bibnamefont {Bouloufa}}, \bibinfo {author} {\bibfnamefont
  {O.}~\bibnamefont {Dulieu}}, \ and\ \bibinfo {author} {\bibfnamefont {H.-C.}\
  \bibnamefont {N{\"a}gerl}},\ }\href {\doibase 10.1039/C1CP21769K} {\bibfield
  {journal} {\bibinfo  {journal} {Phys. Chem. Chem. Phys.}\ }\textbf {\bibinfo
  {volume} {13}},\ \bibinfo {pages} {18926} (\bibinfo {year}
  {2011})}\BibitemShut {NoStop}%
\bibitem [{\citenamefont {Guo}\ \emph {et~al.}(2017)\citenamefont {Guo},
  \citenamefont {Vexiau}, \citenamefont {Zhu}, \citenamefont {Lu},
  \citenamefont {Bouloufa-Maafa}, \citenamefont {Dulieu},\ and\ \citenamefont
  {Wang}}]{Guo2017}%
  \BibitemOpen
  \bibfield  {author} {\bibinfo {author} {\bibfnamefont {M.}~\bibnamefont
  {Guo}}, \bibinfo {author} {\bibfnamefont {R.}~\bibnamefont {Vexiau}},
  \bibinfo {author} {\bibfnamefont {B.}~\bibnamefont {Zhu}}, \bibinfo {author}
  {\bibfnamefont {B.}~\bibnamefont {Lu}}, \bibinfo {author} {\bibfnamefont
  {N.}~\bibnamefont {Bouloufa-Maafa}}, \bibinfo {author} {\bibfnamefont
  {O.}~\bibnamefont {Dulieu}}, \ and\ \bibinfo {author} {\bibfnamefont
  {D.}~\bibnamefont {Wang}},\ }\href {\doibase 10.1103/physreva.96.052505}
  {\bibfield  {journal} {\bibinfo  {journal} {Phys. Rev. A}\ }\textbf {\bibinfo
  {volume} {96}},\ \bibinfo {pages} {052505} (\bibinfo {year}
  {2017})}\BibitemShut {NoStop}%
\bibitem [{\citenamefont {Dutta}\ \emph {et~al.}(2017)\citenamefont {Dutta},
  \citenamefont {P{\'{e}}rez-R{\'{\i}}os}, \citenamefont {Elliott},\ and\
  \citenamefont {Chen}}]{Dutta2017}%
  \BibitemOpen
  \bibfield  {author} {\bibinfo {author} {\bibfnamefont {S.}~\bibnamefont
  {Dutta}}, \bibinfo {author} {\bibfnamefont {J.}~\bibnamefont
  {P{\'{e}}rez-R{\'{\i}}os}}, \bibinfo {author} {\bibfnamefont {D.~S.}\
  \bibnamefont {Elliott}}, \ and\ \bibinfo {author} {\bibfnamefont {Y.~P.}\
  \bibnamefont {Chen}},\ }\href {\doibase 10.1103/physreva.95.013405}
  {\bibfield  {journal} {\bibinfo  {journal} {Phys. Rev. A}\ }\textbf {\bibinfo
  {volume} {95}},\ \bibinfo {pages} {013405} (\bibinfo {year}
  {2017})}\BibitemShut {NoStop}%
\bibitem [{\citenamefont {Rvachov}\ \emph {et~al.}(2018)\citenamefont
  {Rvachov}, \citenamefont {Son}, \citenamefont {Park}, \citenamefont {Ebadi},
  \citenamefont {Zwierlein}, \citenamefont {Ketterle},\ and\ \citenamefont
  {Jamison}}]{Rvachov2018}%
  \BibitemOpen
  \bibfield  {author} {\bibinfo {author} {\bibfnamefont {T.~M.}\ \bibnamefont
  {Rvachov}}, \bibinfo {author} {\bibfnamefont {H.}~\bibnamefont {Son}},
  \bibinfo {author} {\bibfnamefont {J.~J.}\ \bibnamefont {Park}}, \bibinfo
  {author} {\bibfnamefont {S.}~\bibnamefont {Ebadi}}, \bibinfo {author}
  {\bibfnamefont {M.~W.}\ \bibnamefont {Zwierlein}}, \bibinfo {author}
  {\bibfnamefont {W.}~\bibnamefont {Ketterle}}, \ and\ \bibinfo {author}
  {\bibfnamefont {A.~O.}\ \bibnamefont {Jamison}},\ }\href {\doibase
  10.1039/c7cp08481a} {\bibfield  {journal} {\bibinfo  {journal} {Phys. Chem.
  Chem. Phys.}\ }\textbf {\bibinfo {volume} {20}},\ \bibinfo {pages} {4739}
  (\bibinfo {year} {2018})}\BibitemShut {NoStop}%
\bibitem [{\citenamefont {Guttridge}\ \emph {et~al.}(2018)\citenamefont
  {Guttridge}, \citenamefont {Hopkins}, \citenamefont {Frye}, \citenamefont
  {McFerran}, \citenamefont {Hutson},\ and\ \citenamefont
  {Cornish}}]{Guttridge2018}%
  \BibitemOpen
  \bibfield  {author} {\bibinfo {author} {\bibfnamefont {A.}~\bibnamefont
  {Guttridge}}, \bibinfo {author} {\bibfnamefont {S.~A.}\ \bibnamefont
  {Hopkins}}, \bibinfo {author} {\bibfnamefont {M.~D.}\ \bibnamefont {Frye}},
  \bibinfo {author} {\bibfnamefont {J.~J.}\ \bibnamefont {McFerran}}, \bibinfo
  {author} {\bibfnamefont {J.~M.}\ \bibnamefont {Hutson}}, \ and\ \bibinfo
  {author} {\bibfnamefont {S.~L.}\ \bibnamefont {Cornish}},\ }\href {\doibase
  10.1103/PhysRevA.97.063414} {\bibfield  {journal} {\bibinfo  {journal} {Phys.
  Rev. A}\ }\textbf {\bibinfo {volume} {97}},\ \bibinfo {pages} {063414}
  (\bibinfo {year} {2018})}\BibitemShut {NoStop}%
\bibitem [{\citenamefont {Drever}\ \emph {et~al.}(1983)\citenamefont {Drever},
  \citenamefont {Hall}, \citenamefont {Kowalski}, \citenamefont {Hough},
  \citenamefont {Ford}, \citenamefont {Munley},\ and\ \citenamefont
  {Ward}}]{Drever1983}%
  \BibitemOpen
  \bibfield  {author} {\bibinfo {author} {\bibfnamefont {R.~W.~P.}\
  \bibnamefont {Drever}}, \bibinfo {author} {\bibfnamefont {J.~L.}\
  \bibnamefont {Hall}}, \bibinfo {author} {\bibfnamefont {F.~V.}\ \bibnamefont
  {Kowalski}}, \bibinfo {author} {\bibfnamefont {J.}~\bibnamefont {Hough}},
  \bibinfo {author} {\bibfnamefont {G.~M.}\ \bibnamefont {Ford}}, \bibinfo
  {author} {\bibfnamefont {A.~J.}\ \bibnamefont {Munley}}, \ and\ \bibinfo
  {author} {\bibfnamefont {H.}~\bibnamefont {Ward}},\ }\href {\doibase
  10.1007/BF00702605} {\bibfield  {journal} {\bibinfo  {journal} {Appl. Phys.
  B}\ }\textbf {\bibinfo {volume} {31}},\ \bibinfo {pages} {97} (\bibinfo
  {year} {1983})}\BibitemShut {NoStop}%
\bibitem [{\citenamefont {Thorpe}\ \emph {et~al.}(2008)\citenamefont {Thorpe},
  \citenamefont {Numata},\ and\ \citenamefont {Livas}}]{Thorpe2008}%
  \BibitemOpen
  \bibfield  {author} {\bibinfo {author} {\bibfnamefont {J.~I.}\ \bibnamefont
  {Thorpe}}, \bibinfo {author} {\bibfnamefont {K.}~\bibnamefont {Numata}}, \
  and\ \bibinfo {author} {\bibfnamefont {J.}~\bibnamefont {Livas}},\ }\href
  {\doibase 10.1364/oe.16.015980} {\bibfield  {journal} {\bibinfo  {journal}
  {Opt. Express}\ }\textbf {\bibinfo {volume} {16}},\ \bibinfo {pages} {15980}
  (\bibinfo {year} {2008})}\BibitemShut {NoStop}%
\bibitem [{\citenamefont {Gregory}\ \emph {et~al.}(2015)\citenamefont
  {Gregory}, \citenamefont {Molony}, \citenamefont {K\"{o}ppinger},
  \citenamefont {Kumar}, \citenamefont {Ji}, \citenamefont {Lu}, \citenamefont
  {Marchant},\ and\ \citenamefont {Cornish}}]{Gregory2015}%
  \BibitemOpen
  \bibfield  {author} {\bibinfo {author} {\bibfnamefont {P.~D.}\ \bibnamefont
  {Gregory}}, \bibinfo {author} {\bibfnamefont {P.~K.}\ \bibnamefont {Molony}},
  \bibinfo {author} {\bibfnamefont {M.~P.}\ \bibnamefont {K\"{o}ppinger}},
  \bibinfo {author} {\bibfnamefont {A.}~\bibnamefont {Kumar}}, \bibinfo
  {author} {\bibfnamefont {Z.}~\bibnamefont {Ji}}, \bibinfo {author}
  {\bibfnamefont {B.}~\bibnamefont {Lu}}, \bibinfo {author} {\bibfnamefont
  {A.~L.}\ \bibnamefont {Marchant}}, \ and\ \bibinfo {author} {\bibfnamefont
  {S.~L.}\ \bibnamefont {Cornish}},\ }\href {\doibase
  10.1088/1367-2630/17/5/055006} {\bibfield  {journal} {\bibinfo  {journal}
  {New J. Phys.}\ }\textbf {\bibinfo {volume} {17}},\ \bibinfo {pages} {055006}
  (\bibinfo {year} {2015})}\BibitemShut {NoStop}%
\bibitem [{\citenamefont {Winkler}\ \emph {et~al.}(2005)\citenamefont
  {Winkler}, \citenamefont {Thalhammer}, \citenamefont {Theis}, \citenamefont
  {Ritsch}, \citenamefont {Grimm},\ and\ \citenamefont {{Hecker
  Denschlag}}}]{Winkler2005}%
  \BibitemOpen
  \bibfield  {author} {\bibinfo {author} {\bibfnamefont {K.}~\bibnamefont
  {Winkler}}, \bibinfo {author} {\bibfnamefont {G.}~\bibnamefont {Thalhammer}},
  \bibinfo {author} {\bibfnamefont {M.}~\bibnamefont {Theis}}, \bibinfo
  {author} {\bibfnamefont {H.}~\bibnamefont {Ritsch}}, \bibinfo {author}
  {\bibfnamefont {R.}~\bibnamefont {Grimm}}, \ and\ \bibinfo {author}
  {\bibfnamefont {J.}~\bibnamefont {{Hecker Denschlag}}},\ }\href {\doibase
  10.1103/PhysRevLett.95.063202} {\bibfield  {journal} {\bibinfo  {journal}
  {Phys. Rev. Lett.}\ }\textbf {\bibinfo {volume} {95}},\ \bibinfo {pages}
  {063202} (\bibinfo {year} {2005})}\BibitemShut {NoStop}%
\bibitem [{\citenamefont {Fleischhauer}\ \emph {et~al.}(2005)\citenamefont
  {Fleischhauer}, \citenamefont {Imamoglu},\ and\ \citenamefont
  {Marangos}}]{Fleischhauer2005}%
  \BibitemOpen
  \bibfield  {author} {\bibinfo {author} {\bibfnamefont {M.}~\bibnamefont
  {Fleischhauer}}, \bibinfo {author} {\bibfnamefont {A.}~\bibnamefont
  {Imamoglu}}, \ and\ \bibinfo {author} {\bibfnamefont {J.~P.}\ \bibnamefont
  {Marangos}},\ }\href {\doibase 10.1103/revmodphys.77.633} {\bibfield
  {journal} {\bibinfo  {journal} {Rev. Mod. Phys.}\ }\textbf {\bibinfo {volume}
  {77}},\ \bibinfo {pages} {633} (\bibinfo {year} {2005})}\BibitemShut
  {NoStop}%
\bibitem [{Note1()}]{Note1}%
  \BibitemOpen
  \bibinfo {note} {The production of Cs*Yb molecules causes a detectable loss
  of Cs atoms from the trap}\BibitemShut {NoStop}%
\bibitem [{\citenamefont {Bohn}\ and\ \citenamefont
  {Julienne}(1996)}]{Bohn1996}%
  \BibitemOpen
  \bibfield  {author} {\bibinfo {author} {\bibfnamefont {J.~L.}\ \bibnamefont
  {Bohn}}\ and\ \bibinfo {author} {\bibfnamefont {P.~S.}\ \bibnamefont
  {Julienne}},\ }\href {\doibase 10.1103/physreva.54.r4637} {\bibfield
  {journal} {\bibinfo  {journal} {Phys. Rev. A}\ }\textbf {\bibinfo {volume}
  {54}},\ \bibinfo {pages} {R4637} (\bibinfo {year} {1996})}\BibitemShut
  {NoStop}%
\bibitem [{\citenamefont {Portier}\ \emph {et~al.}(2009)\citenamefont
  {Portier}, \citenamefont {Leduc},\ and\ \citenamefont
  {Cohen-Tannoudji}}]{Portier2009}%
  \BibitemOpen
  \bibfield  {author} {\bibinfo {author} {\bibfnamefont {M.}~\bibnamefont
  {Portier}}, \bibinfo {author} {\bibfnamefont {M.}~\bibnamefont {Leduc}}, \
  and\ \bibinfo {author} {\bibfnamefont {C.}~\bibnamefont {Cohen-Tannoudji}},\
  }\href {\doibase 10.1039/b819470j} {\bibfield  {journal} {\bibinfo  {journal}
  {Faraday Discuss.}\ }\textbf {\bibinfo {volume} {142}},\ \bibinfo {pages}
  {415} (\bibinfo {year} {2009})}\BibitemShut {NoStop}%
\bibitem [{\citenamefont {Fano}(1961)}]{Fano1961}%
  \BibitemOpen
  \bibfield  {author} {\bibinfo {author} {\bibfnamefont {U.}~\bibnamefont
  {Fano}},\ }\href {\doibase 10.1103/physrev.124.1866} {\bibfield  {journal}
  {\bibinfo  {journal} {Phys. Rev.}\ }\textbf {\bibinfo {volume} {124}},\
  \bibinfo {pages} {1866} (\bibinfo {year} {1961})}\BibitemShut {NoStop}%
\bibitem [{\citenamefont {Brewer}\ and\ \citenamefont
  {Hahn}(1975)}]{Brewer1975}%
  \BibitemOpen
  \bibfield  {author} {\bibinfo {author} {\bibfnamefont {R.~G.}\ \bibnamefont
  {Brewer}}\ and\ \bibinfo {author} {\bibfnamefont {E.~L.}\ \bibnamefont
  {Hahn}},\ }\href {\doibase 10.1103/physreva.11.1641} {\bibfield  {journal}
  {\bibinfo  {journal} {Phys. Rev. A}\ }\textbf {\bibinfo {volume} {11}},\
  \bibinfo {pages} {1641} (\bibinfo {year} {1975})}\BibitemShut {NoStop}%
\bibitem [{\citenamefont {Orriols}(1979)}]{Orriols1979}%
  \BibitemOpen
  \bibfield  {author} {\bibinfo {author} {\bibfnamefont {G.}~\bibnamefont
  {Orriols}},\ }\href {\doibase 10.1007/bf02739299} {\bibfield  {journal}
  {\bibinfo  {journal} {Nuovo Cimento B}\ }\textbf {\bibinfo {volume} {53}},\
  \bibinfo {pages} {1} (\bibinfo {year} {1979})}\BibitemShut {NoStop}%
\bibitem [{\citenamefont {Lounis}\ and\ \citenamefont
  {Cohen-Tannoudji}(1992)}]{Lounis1992}%
  \BibitemOpen
  \bibfield  {author} {\bibinfo {author} {\bibfnamefont {B.}~\bibnamefont
  {Lounis}}\ and\ \bibinfo {author} {\bibfnamefont {C.}~\bibnamefont
  {Cohen-Tannoudji}},\ }\href {\doibase 10.1051/jp2:1992153} {\bibfield
  {journal} {\bibinfo  {journal} {J. Phys.}\ }\textbf {\bibinfo {volume} {2}},\
  \bibinfo {pages} {579} (\bibinfo {year} {1992})}\BibitemShut {NoStop}%
\bibitem [{\citenamefont {Zanon-Willette}\ \emph {et~al.}(2011)\citenamefont
  {Zanon-Willette}, \citenamefont {de~Clercq},\ and\ \citenamefont
  {Arimondo}}]{Zanon-Willette2011}%
  \BibitemOpen
  \bibfield  {author} {\bibinfo {author} {\bibfnamefont {T.}~\bibnamefont
  {Zanon-Willette}}, \bibinfo {author} {\bibfnamefont {E.}~\bibnamefont
  {de~Clercq}}, \ and\ \bibinfo {author} {\bibfnamefont {E.}~\bibnamefont
  {Arimondo}},\ }\href {\doibase 10.1103/physreva.84.062502} {\bibfield
  {journal} {\bibinfo  {journal} {Phys. Rev. A}\ }\textbf {\bibinfo {volume}
  {84}},\ \bibinfo {pages} {062502} (\bibinfo {year} {2011})}\BibitemShut
  {NoStop}%
\bibitem [{\citenamefont {Cohen-Tannoudji}(2015)}]{Cohen-Tannoudji2015}%
  \BibitemOpen
  \bibfield  {author} {\bibinfo {author} {\bibfnamefont {C.}~\bibnamefont
  {Cohen-Tannoudji}},\ }\href {\doibase 10.1088/0031-8949/90/8/088013}
  {\bibfield  {journal} {\bibinfo  {journal} {Phys. Scr.}\ }\textbf {\bibinfo
  {volume} {90}},\ \bibinfo {pages} {088013} (\bibinfo {year}
  {2015})}\BibitemShut {NoStop}%
\bibitem [{Note2()}]{Note2}%
  \BibitemOpen
  \bibinfo {note} {The definition of $\Omega _{\protect \rm BB}$ in Eq.~\ref
  {eq:shift} is twice that in Ref.~\cite {Portier2009}}\BibitemShut {NoStop}%
\bibitem [{\citenamefont {{Le Roy}}\ and\ \citenamefont
  {Bernstein}(1970)}]{LeRoy1970}%
  \BibitemOpen
  \bibfield  {author} {\bibinfo {author} {\bibfnamefont {R.~J.}\ \bibnamefont
  {{Le Roy}}}\ and\ \bibinfo {author} {\bibfnamefont {R.~B.}\ \bibnamefont
  {Bernstein}},\ }\href {\doibase 10.1063/1.1673585} {\bibfield  {journal}
  {\bibinfo  {journal} {J. Chem. Phys.}\ }\textbf {\bibinfo {volume} {52}},\
  \bibinfo {pages} {3869} (\bibinfo {year} {1970})}\BibitemShut {NoStop}%
\bibitem [{\citenamefont {Meyer}\ and\ \citenamefont {Bohn}(2009)}]{Meyer2009}%
  \BibitemOpen
  \bibfield  {author} {\bibinfo {author} {\bibfnamefont {E.~R.}\ \bibnamefont
  {Meyer}}\ and\ \bibinfo {author} {\bibfnamefont {J.~L.}\ \bibnamefont
  {Bohn}},\ }\href {\doibase 10.1103/PhysRevA.80.042508} {\bibfield  {journal}
  {\bibinfo  {journal} {Phys. Rev. A}\ }\textbf {\bibinfo {volume} {80}},\
  \bibinfo {pages} {042508} (\bibinfo {year} {2009})}\BibitemShut {NoStop}%
\bibitem [{\citenamefont {Shao}\ \emph {et~al.}(2017)\citenamefont {Shao},
  \citenamefont {Deng}, \citenamefont {Xing}, \citenamefont {Gou},
  \citenamefont {Kuang},\ and\ \citenamefont {Li}}]{Shao2017}%
  \BibitemOpen
  \bibfield  {author} {\bibinfo {author} {\bibfnamefont {Q.}~\bibnamefont
  {Shao}}, \bibinfo {author} {\bibfnamefont {L.}~\bibnamefont {Deng}}, \bibinfo
  {author} {\bibfnamefont {X.}~\bibnamefont {Xing}}, \bibinfo {author}
  {\bibfnamefont {D.}~\bibnamefont {Gou}}, \bibinfo {author} {\bibfnamefont
  {X.}~\bibnamefont {Kuang}}, \ and\ \bibinfo {author} {\bibfnamefont
  {H.}~\bibnamefont {Li}},\ }\href {\doibase 10.1021/acs.jpca.6b11741}
  {\bibfield  {journal} {\bibinfo  {journal} {J. Phys. Chem. A}\ }\textbf
  {\bibinfo {volume} {121}},\ \bibinfo {pages} {2187} (\bibinfo {year}
  {2017})}\BibitemShut {NoStop}%
\bibitem [{Note3()}]{Note3}%
  \BibitemOpen
  \bibinfo {note} {3.9 eV for Cs and 6.3 eV for Yb.}\BibitemShut {Stop}%
\bibitem [{\citenamefont {Tang}\ and\ \citenamefont
  {Toennies}(1984)}]{Tang1984}%
  \BibitemOpen
  \bibfield  {author} {\bibinfo {author} {\bibfnamefont {K.~T.}\ \bibnamefont
  {Tang}}\ and\ \bibinfo {author} {\bibfnamefont {J.~P.}\ \bibnamefont
  {Toennies}},\ }\href {\doibase 10.1063/1.447150} {\bibfield  {journal}
  {\bibinfo  {journal} {J. Chem. Phys.}\ }\textbf {\bibinfo {volume} {80}},\
  \bibinfo {pages} {3726} (\bibinfo {year} {1984})}\BibitemShut {NoStop}%
\bibitem [{\citenamefont {Thakkar}\ and\ \citenamefont
  {Smith}(1974)}]{Thakkar1974}%
  \BibitemOpen
  \bibfield  {author} {\bibinfo {author} {\bibfnamefont {A.~J.}\ \bibnamefont
  {Thakkar}}\ and\ \bibinfo {author} {\bibfnamefont {V.~H.}\ \bibnamefont
  {Smith}},\ }\href {\doibase 10.1088/0022-3700/7/10/004} {\bibfield  {journal}
  {\bibinfo  {journal} {J. Phys. B: At. Mol. Phys.}\ }\textbf {\bibinfo
  {volume} {7}},\ \bibinfo {pages} {L321} (\bibinfo {year} {1974})}\BibitemShut
  {NoStop}%
\bibitem [{\citenamefont {Tang}(1969)}]{Tang1969}%
  \BibitemOpen
  \bibfield  {author} {\bibinfo {author} {\bibfnamefont {K.~T.}\ \bibnamefont
  {Tang}},\ }\href {\doibase 10.1103/physrev.177.108} {\bibfield  {journal}
  {\bibinfo  {journal} {Phys. Rev.}\ }\textbf {\bibinfo {volume} {177}},\
  \bibinfo {pages} {108} (\bibinfo {year} {1969})}\BibitemShut {NoStop}%
\bibitem [{\citenamefont {{Le Roy}}(1998)}]{leRoy1998}%
  \BibitemOpen
  \bibfield  {author} {\bibinfo {author} {\bibfnamefont {R.~J.}\ \bibnamefont
  {{Le Roy}}},\ }\href {\doibase 10.1006/jmsp.1998.7646} {\bibfield  {journal}
  {\bibinfo  {journal} {J. Mol. Spectrosc.}\ }\textbf {\bibinfo {volume}
  {191}},\ \bibinfo {pages} {223} (\bibinfo {year} {1998})}\BibitemShut
  {NoStop}%
\bibitem [{\citenamefont {Hutson}(1993)}]{Hutson:bound:1993}%
  \BibitemOpen
  \bibfield  {author} {\bibinfo {author} {\bibfnamefont {J.~M.}\ \bibnamefont
  {Hutson}},\ }\href@noop {} {\enquote {\bibinfo {title} {{BOUND} computer
  program, version 5},}\ }\bibinfo {howpublished} {Distributed by Collaborative
  Computational Project No.\ 6 of the UK Engineering and Physical Sciences
  Research Council} (\bibinfo {year} {1993})\BibitemShut {NoStop}%
\bibitem [{Note4()}]{Note4}%
  \BibitemOpen
  \bibinfo {note} {In principle, the potential might support different numbers
  of vibrational levels for different isotopologs, but we find that the three
  isotopologs for which we have measurements have the same number of
  vibrational levels.}\BibitemShut {Stop}%
\bibitem [{\citenamefont {Gribakin}\ and\ \citenamefont
  {Flambaum}(1993)}]{Gribakin1993}%
  \BibitemOpen
  \bibfield  {author} {\bibinfo {author} {\bibfnamefont {G.~F.}\ \bibnamefont
  {Gribakin}}\ and\ \bibinfo {author} {\bibfnamefont {V.~V.}\ \bibnamefont
  {Flambaum}},\ }\href {\doibase https://doi.org/10.1103/PhysRevA.48.546}
  {\bibfield  {journal} {\bibinfo  {journal} {Phys. Rev. A}\ }\textbf {\bibinfo
  {volume} {48}},\ \bibinfo {pages} {546} (\bibinfo {year} {1993})}\BibitemShut
  {NoStop}%
\bibitem [{\citenamefont {Riboli}\ and\ \citenamefont
  {Modugno}(2002)}]{Riboli2002}%
  \BibitemOpen
  \bibfield  {author} {\bibinfo {author} {\bibfnamefont {F.}~\bibnamefont
  {Riboli}}\ and\ \bibinfo {author} {\bibfnamefont {M.}~\bibnamefont
  {Modugno}},\ }\href {\doibase 10.1103/PhysRevA.65.063614} {\bibfield
  {journal} {\bibinfo  {journal} {Phys. Rev. A}\ }\textbf {\bibinfo {volume}
  {65}},\ \bibinfo {pages} {063614} (\bibinfo {year} {2002})}\BibitemShut
  {NoStop}%
\bibitem [{\citenamefont {Weber}\ \emph {et~al.}(2003)\citenamefont {Weber},
  \citenamefont {Herbig}, \citenamefont {Mark}, \citenamefont {N\"agerl},\ and\
  \citenamefont {Grimm}}]{Weber2003}%
  \BibitemOpen
  \bibfield  {author} {\bibinfo {author} {\bibfnamefont {T.}~\bibnamefont
  {Weber}}, \bibinfo {author} {\bibfnamefont {J.}~\bibnamefont {Herbig}},
  \bibinfo {author} {\bibfnamefont {M.}~\bibnamefont {Mark}}, \bibinfo {author}
  {\bibfnamefont {H.-C.}\ \bibnamefont {N\"agerl}}, \ and\ \bibinfo {author}
  {\bibfnamefont {R.}~\bibnamefont {Grimm}},\ }\href {\doibase
  10.1103/PhysRevLett.91.123201} {\bibfield  {journal} {\bibinfo  {journal}
  {Phys. Rev. Lett.}\ }\textbf {\bibinfo {volume} {91}},\ \bibinfo {pages}
  {123201} (\bibinfo {year} {2003})}\BibitemShut {NoStop}%
\bibitem [{\citenamefont {Petrov}(2015)}]{Petrov2015}%
  \BibitemOpen
  \bibfield  {author} {\bibinfo {author} {\bibfnamefont {D.~S.}\ \bibnamefont
  {Petrov}},\ }\href {\doibase 10.1103/PhysRevLett.115.155302} {\bibfield
  {journal} {\bibinfo  {journal} {Phys. Rev. Lett.}\ }\textbf {\bibinfo
  {volume} {115}},\ \bibinfo {pages} {155302} (\bibinfo {year}
  {2015})}\BibitemShut {NoStop}%
\bibitem [{\citenamefont {Cabrera}\ \emph {et~al.}(2018)\citenamefont
  {Cabrera}, \citenamefont {Tanzi}, \citenamefont {Sanz}, \citenamefont
  {Naylor}, \citenamefont {Thomas}, \citenamefont {Cheiney},\ and\
  \citenamefont {Tarruell}}]{Cabrera2018}%
  \BibitemOpen
  \bibfield  {author} {\bibinfo {author} {\bibfnamefont {C.~R.}\ \bibnamefont
  {Cabrera}}, \bibinfo {author} {\bibfnamefont {L.}~\bibnamefont {Tanzi}},
  \bibinfo {author} {\bibfnamefont {J.}~\bibnamefont {Sanz}}, \bibinfo {author}
  {\bibfnamefont {B.}~\bibnamefont {Naylor}}, \bibinfo {author} {\bibfnamefont
  {P.}~\bibnamefont {Thomas}}, \bibinfo {author} {\bibfnamefont
  {P.}~\bibnamefont {Cheiney}}, \ and\ \bibinfo {author} {\bibfnamefont
  {L.}~\bibnamefont {Tarruell}},\ }\href {\doibase 10.1126/science.aao5686}
  {\bibfield  {journal} {\bibinfo  {journal} {Science}\ }\textbf {\bibinfo
  {volume} {359}},\ \bibinfo {pages} {301} (\bibinfo {year}
  {2018})}\BibitemShut {NoStop}%
\bibitem [{\citenamefont {Cheiney}\ \emph {et~al.}(2018)\citenamefont
  {Cheiney}, \citenamefont {Cabrera}, \citenamefont {Sanz}, \citenamefont
  {Naylor}, \citenamefont {Tanzi},\ and\ \citenamefont
  {Tarruell}}]{Cheiney2018}%
  \BibitemOpen
  \bibfield  {author} {\bibinfo {author} {\bibfnamefont {P.}~\bibnamefont
  {Cheiney}}, \bibinfo {author} {\bibfnamefont {C.~R.}\ \bibnamefont
  {Cabrera}}, \bibinfo {author} {\bibfnamefont {J.}~\bibnamefont {Sanz}},
  \bibinfo {author} {\bibfnamefont {B.}~\bibnamefont {Naylor}}, \bibinfo
  {author} {\bibfnamefont {L.}~\bibnamefont {Tanzi}}, \ and\ \bibinfo {author}
  {\bibfnamefont {L.}~\bibnamefont {Tarruell}},\ }\href {\doibase
  10.1103/PhysRevLett.120.135301} {\bibfield  {journal} {\bibinfo  {journal}
  {Phys. Rev. Lett.}\ }\textbf {\bibinfo {volume} {120}},\ \bibinfo {pages}
  {135301} (\bibinfo {year} {2018})}\BibitemShut {NoStop}%
\bibitem [{\citenamefont {Frye}\ and\ \citenamefont {Hutson}(2014)}]{Frye2014}%
  \BibitemOpen
  \bibfield  {author} {\bibinfo {author} {\bibfnamefont {M.~D.}\ \bibnamefont
  {Frye}}\ and\ \bibinfo {author} {\bibfnamefont {J.~M.}\ \bibnamefont
  {Hutson}},\ }\href {\doibase 10.1103/PhysRevA.89.052705} {\bibfield
  {journal} {\bibinfo  {journal} {Phys. Rev. A}\ }\textbf {\bibinfo {volume}
  {89}},\ \bibinfo {pages} {052705} (\bibinfo {year} {2014})}\BibitemShut
  {NoStop}%
\bibitem [{\citenamefont {Hutson}\ and\ \citenamefont
  {Green}(1994)}]{molscat:v14}%
  \BibitemOpen
  \bibfield  {author} {\bibinfo {author} {\bibfnamefont {J.~M.}\ \bibnamefont
  {Hutson}}\ and\ \bibinfo {author} {\bibfnamefont {S.}~\bibnamefont {Green}},\
  }\href@noop {} {\enquote {\bibinfo {title} {{MOLSCAT} computer program,
  version 14},}\ }\bibinfo {howpublished} {distributed by Collaborative
  Computational Project No.\ 6 of the UK Engineering and Physical Sciences
  Research Council} (\bibinfo {year} {1994})\BibitemShut {NoStop}%
\bibitem [{\citenamefont {Hutson}\ and\ \citenamefont {Green}(1982)}]{sbe}%
  \BibitemOpen
  \bibfield  {author} {\bibinfo {author} {\bibfnamefont {J.~M.}\ \bibnamefont
  {Hutson}}\ and\ \bibinfo {author} {\bibfnamefont {S.}~\bibnamefont {Green}},\
  }\href@noop {} {\enquote {\bibinfo {title} {{SBE} computer program},}\
  }\bibinfo {howpublished} {distributed by Collaborative Computational Project
  No.\ 6 of the UK Engineering and Physical Sciences Research Council}
  (\bibinfo {year} {1982})\BibitemShut {NoStop}%
\bibitem [{\citenamefont {Bergmann}\ \emph {et~al.}(1998)\citenamefont
  {Bergmann}, \citenamefont {Theuer},\ and\ \citenamefont
  {Shore}}]{Bergmann1998}%
  \BibitemOpen
  \bibfield  {author} {\bibinfo {author} {\bibfnamefont {K.}~\bibnamefont
  {Bergmann}}, \bibinfo {author} {\bibfnamefont {H.}~\bibnamefont {Theuer}}, \
  and\ \bibinfo {author} {\bibfnamefont {B.}~\bibnamefont {Shore}},\ }\href
  {\doibase 10.1103/RevModPhys.70.1003} {\bibfield  {journal} {\bibinfo
  {journal} {Rev. Mod. Phys.}\ }\textbf {\bibinfo {volume} {70}},\ \bibinfo
  {pages} {1003} (\bibinfo {year} {1998})}\BibitemShut {NoStop}%
\end{thebibliography}%

\end{document}